\documentclass[a4paper]{aa}

\usepackage{graphicx,natbib}
\usepackage{txfonts}
\usepackage{color}
\usepackage[usenames,dvipsnames]{xcolor}

\newcommand{\bz}{$\langle B_\mathrm{z} \rangle$}
\newcommand{\uma}{$\varepsilon$~UMa}
\newcommand{\aur}{$\theta$~Aur}
\newcommand{\teff}{$T_{\rm eff}$}
\newcommand{\lgg}{$\log g$}
\newcommand{\vs}{$v_{\rm e}\sin i$}
\newcommand{\kms}{km\,s$^{-1}$}

\newcommand{\figps}[1]{\resizebox{\hsize}{!}{\rotatebox{0}{\includegraphics{#1}}}}

\newcommand{\fups}[3]{\resizebox{#1}{!}{\rotatebox{#2}{\includegraphics{#3}}}}

\newcommand{\beq}{\begin{equation}}
\newcommand{\eeq}{\end{equation}}

\begin{document}

\title{Magnetic field topologies of the bright, weak-field Ap stars \\ $\theta$~Aurigae and $\varepsilon$~Ursae Majoris}

\author{O.~Kochukhov\inst{1}
   \and M.~Shultz\inst{1}\thanks{Current address: Department of Physics and Astronomy, University of Delaware, Newark, DE 19711, USA}
   \and C.~Neiner\inst{2}}

\institute{
Department of Physics and Astronomy, Uppsala University, Box 516, SE-75120 Uppsala, Sweden\\\email{oleg.kochukhov@physics.uu.se}
\and
LESIA, Observatoire de Paris, PSL Research University, CNRS, Sorbonne Universit\'{e},
Univ. Paris Diderot, Sorbonne Paris Cit\'{e}, 5 place Jules Janssen, F-92195 Meudon, France
}

\date{Received 19 September 2018 / Accepted 12 November 2018}

\titlerunning{Magnetic field topologies of $\theta$~Aur and $\varepsilon$~UMa}
\authorrunning{O.~Kochukhov et al.}

\abstract
%context
{
The brightest magnetic chemically peculiar stars \aur\ and \uma\ were targeted by numerous studies of their photometric and spectroscopic variability. Detailed maps of chemical abundance spots were repeatedly derived for both stars. However, owing to the weakness of their surface magnetic fields, very little information on the magnetic field geometries of these stars is available.
}
%aims
{
In this study we aim to determine detailed magnetic field topologies of \aur\ and \uma\ based on modern, high-resolution spectropolarimetric observations. 
}
%methods
{
Both targets were observed in all four Stokes parameters using the Narval and ESPaDOnS spectropolarimeters. A multi-line technique of least-squares deconvolution was employed to detect polarisation signatures in spectral lines. These signatures were modelled with a Zeeman-Doppler imaging code. 
}
%results
{
We succeeded in detecting variable circular and linear polarisation signatures for \aur. Only circular polarisation was detected for \uma. We obtained new sets of high-precision longitudinal magnetic field measurements using mean circular polarisation metal line profiles as well as hydrogen line cores, which are consistent with historical data. Magnetic inversions revealed distorted dipolar geometries in both stars. The Fe and Cr abundance distributions, reconstructed simultaneously with magnetic mapping, do not show a clear correlation with the local magnetic field properties, with the exception of a relative element underabundance in the horizontal field regions along the magnetic equators.
}
%conclusions
{
Our study provides the first ever detailed surface magnetic field maps for broad-line, weak-field chemically peculiar stars, showing that their field topologies are qualitatively similar to those found in stronger-field stars. The Fe and Cr chemical abundance maps reconstructed for \aur\ and \uma\ are at odds with the predictions of current theoretical atomic diffusion calculations.
}


\maketitle

\section{Introduction}
\label{intro}

Investigation of stellar magnetic fields and related surface activity and structure formation processes is one of the key research directions of modern stellar physics. In this context, the upper main sequence, magnetic chemically peculiar (MCP/ApBp) stars offer particularly attractive natural laboratories thanks to their strong, globally-organised and stable magnetic fields accompanied by pronounced vertical and horizontal chemical abundance inhomogeneities. Detailed knowledge of the magnetic field geometries of MCP stars is essential for testing theories of the origin and evolution of fossil magnetic fields in stellar interiors \citep{braithwaite:2006,duez:2010} and provides critical constraints for the studies of magnetically-confined stellar winds \citep{babel:1992,babel:1997,ud-doula:2002} and radiatively-driven segregation of chemical elements \citep{leblanc:2009,alecian:2015}.

However, none of the MCP stars are close enough to be readily accessible to a direct surface structure investigation using high-contrast imaging or interferometric techniques \citep{shulyak:2014a}. Instead, the indirect surface mapping methods, such as Doppler imaging (DI) and Zeeman-Doppler imaging (ZDI) \citep{kochukhov:2016}, currently represent the only viable options for reconstructing surface structure maps from the rotational modulation of the intensity and polarisation line profiles. As summarised by \citet{kochukhov:2017}, chemical spot maps have been published for about 40 A and B-type MCP stars. At the same time, detailed ZDI magnetic field maps are available only for about a dozen of these objects. An even smaller number of stars were investigated using high-resolution spectra in all four Stokes parameters, which are essential for the full characterisation of stellar magnetic fields and, in particular, for revealing smaller-scale aspects of the surface field topologies \citep{kochukhov:2004d,kochukhov:2010,kochukhov:2016a}. The vast majority of recent ZDI studies of MCP stars also tend to be biased towards narrow-line stars with stronger than average magnetic fields \citep[e.g.][]{kochukhov:2015,rusomarov:2016,rusomarov:2018,silvester:2014a,silvester:2015,silvester:2017,yakunin:2015}. Relatively little information is available about the structure of the weak magnetic fields of fast-rotating MCP stars, in spite of the fact that these objects include some of the brightest MCP stars with the most complete constraints on fundamental parameters, rotational variability, and surface chemistry. The goal of this paper is to alleviate this observational bias by performing a detailed magnetic field topology analysis of two bright and otherwise very well-studied MCP stars, \aur\ and \uma.

The second brightest MCP star, \aur\ (HD\,40312, HR\,2095, spectral type A0p Si), is a frequent target of photometric, spectroscopic and spectrophotometric variability studies \citep[see][and references therein]{krticka:2015}. It is also one of the first rotationally variable $\alpha^2$~CVn-type stars for which horizontal star spot maps were reconstructed with the DI technique \citep{khokhlova:1986}. More recently, surface mapping of different chemical elements was carried out by \citet{rice:1990}, \citet{hatzes:1991}, \citet{rice:2004}, and \citet{kuschnig:1998}. \citet{krticka:2015} used empirical chemical maps from the latter study to reproduce photometric light curves as well as rotational modulation of the UV stellar flux distribution. 

Relatively little is known about the magnetic field geometry of \aur\ and its relation to the surface abundance inhomogeneities inferred by the DI studies of that star. Measurements of the mean longitudinal magnetic field \bz\ by \citet{landstreet:1977}, \citet{borra:1980}, \citet{wade:2000}, and \citet{silvester:2012} revealed a reversing, nearly sinusoidal variation spanning a range of $\pm300$~G. According to \citet{shulyak:2007}, this longitudinal field curve indicates a dipolar field intensity of 1.3--1.4~kG but provides little constraint on the quadrupolar component. This curve alone cannot be used to investigate magnetic structures at the smaller spatial scales, which are readily resolved by the abundance DI studies of \aur. 

The second target of our study, \uma\ (HD\,112185, HR\,4905, Alioth), is the brightest MCP star in the sky. It is spectrally classified as A1p CrEuMn \citep{renson:2009}. Similar to \aur, variability of \uma\ was investigated during nearly a century of photometric and spectroscopic observations \citep[see][and references therein]{shulyak:2010b}. This star was also frequently targeted by DI studies \citep{rice:1989,rice:1997,lueftinger:2003}. \citet{shulyak:2010b} were able to reproduce most of the observed light variability of \uma\ using multi-element DI maps provided by \citet{lueftinger:2003}.

Conclusive longitudinal magnetic field measurements were made for \uma\ by \citet{bohlender:1990a}, \citet{donati:1990a}, and \citet{wade:2000}. These studies showed that \bz\ varies roughly from $-100$ to $+100$~G. This implies that, despite prominent rotational variability, \uma\ hosts one of the weakest magnetic fields among MCP stars. \citet{shulyak:2010b} inferred a conservative upper limit of 400~G for the dipolar field strength from the \bz\ curve, while \citet{donati:1990a} derived a dipolar field intensity of 186~G from modelling five Stokes $V$ observations of a single magnetically sensitive \ion{Fe}{ii} line. None of the previous studies of \uma\ provided constraints on field components more complex than a dipole.

In this paper we aim to investigate in detail the magnetic field topologies of \aur\ and \uma, thereby probing the previously unexplored regime of rapid rotation and weak field. We describe our new high quality, four Stokes parameter spectropolarimetric observations of these MCP stars in Sect.~\ref{obs}. A revision of stellar fundamental parameters based on the latest generation of individualised stellar atmosphere models is presented in Sect.~\ref{params}. This is followed by the discussion of the longitudinal magnetic field measurements in Sect.~\ref{bz} and ZDI modelling methodology and results in Sect.~\ref{zdi}. The main findings of our study are summarised and discussed in Sect.~\ref{conc}.

\section{Spectropolarimetric observations}
\label{obs}

The bulk of the observations of \aur\ and \uma\ analysed in this paper were obtained with the Narval spectropolarimeter \citep{auriere:2003} in the context of the BritePol observing campaign \citep{neiner:2017}. Narval is  a high-resolution echelle spectropolarimeter, fibre-fed from the Cassegrain focus of the 2-m Bernard Lyot Telescope of the Pic du Midi observatory. This instrument has a resolving power of 65000 and covers a spectral range 370--1050~nm in a single exposure. Narval is capable of obtaining four Stokes parameter (Stokes $IQUV$) spectra employing an efficient beam-switching technique to suppress instrumental polarisation artefacts \citep{donati:1997}. Each polarimetric observation comprises at least four sub-exposures obtained with different configuration of the polarimeter. The resulting pairs of orthogonally polarised spectra are added together to obtain the intensity (Stokes $I$) spectrum and combined according to the ``ratio'' polarimetric demodulation method \citep{bagnulo:2009} to yield one of the polarisation (Stokes $V$, $Q$, $U$) spectra as well as the corresponding diagnostic null spectrum. 

The paper by \citet{silvester:2012} provides a detailed description of the reduction of Narval four Stokes parameter observations. These authors also assess polarimetric accuracy and calibration of this instrument in comparison to the twin ESPaDOnS spectropolarimeter mounted at the Canada-France-Hawaii Telescope (CFHT). The standard automatic reduction software running at the Pic du Midi observatory, identical to that described by \citet{silvester:2012}, was used in our study for all reduction steps except the final continuum normalisation. The latter was performed with a global fit, using the method and routines described by \citet{rosen:2018}.

Our Narval observations of \aur\ were obtained between September 2016 and April 2017. We have secured 19 full four Stokes parameter observations and one $IQU$ observation with typical total exposure times of 960~s for circular polarisation and 1920~s for each of the two linear polarisation parameters. Each of these Stokes parameter observations was split into 4 to 16 sub-exposures, which allowed us to reach a high signal-to-noise ratio (SNR) without saturating the detector. In addition, we made use of one Stokes $V$ and 8 Stokes $QUV$ archival observations acquired with Narval and ESPaDOnS in 2006 and 2008. These data were previously analysed by \citet{silvester:2012}.

The median SNR of all 29 \aur\ observations is 1425 for the Stokes $V$ and 2070 for the Stokes $QU$ parameters. A detailed log of these data is given in Table~\ref{tbl:obs}. The heliocentric Julian dates and the rotational phases reported in this table correspond to the mean values of consecutive Stokes parameter observations. These observing sequences were always obtained during less than 2\% of the stellar rotational period. The rotational ephemeris of \aur,
\beq
HJD = 2450001.881 + 3\fd 618664 \times E,
\eeq
was adopted from the study by \citet{krticka:2015}.

\uma\ was observed with Narval from December 2016 to April 2017. During this period we obtained 26 full Stokes vector observations with a total exposure times 156--312~s per Stokes parameter and a median SNR of 1055 for Stokes $V$ and 1518 for Stokes $QU$. Each Stokes parameter observation was split into 4 or 8 sub-exposures. One additional archival Stokes $V$ spectrum of \uma, acquired with Narval in 2014, was included in the analysis. The log of spectropolarimetric observational data collected for \uma\ is provided in Table~\ref{tbl:obs}. The rotational phases reported in this table were calculated using the ephemeris
\beq
HJD = 2442150.778 + 5\fd088631 \times E
\eeq
derived by \citet{shulyak:2010b}. All consecutive Stokes parameter observations of \uma\ were obtained within no more than 0.3\% of the rotational period, justifying the use of mean heliocentric Julian dates and rotational phases for the line profile analysis performed later in this paper.

\begin{figure*}[!th]
\centering
\fups{16cm}{90}{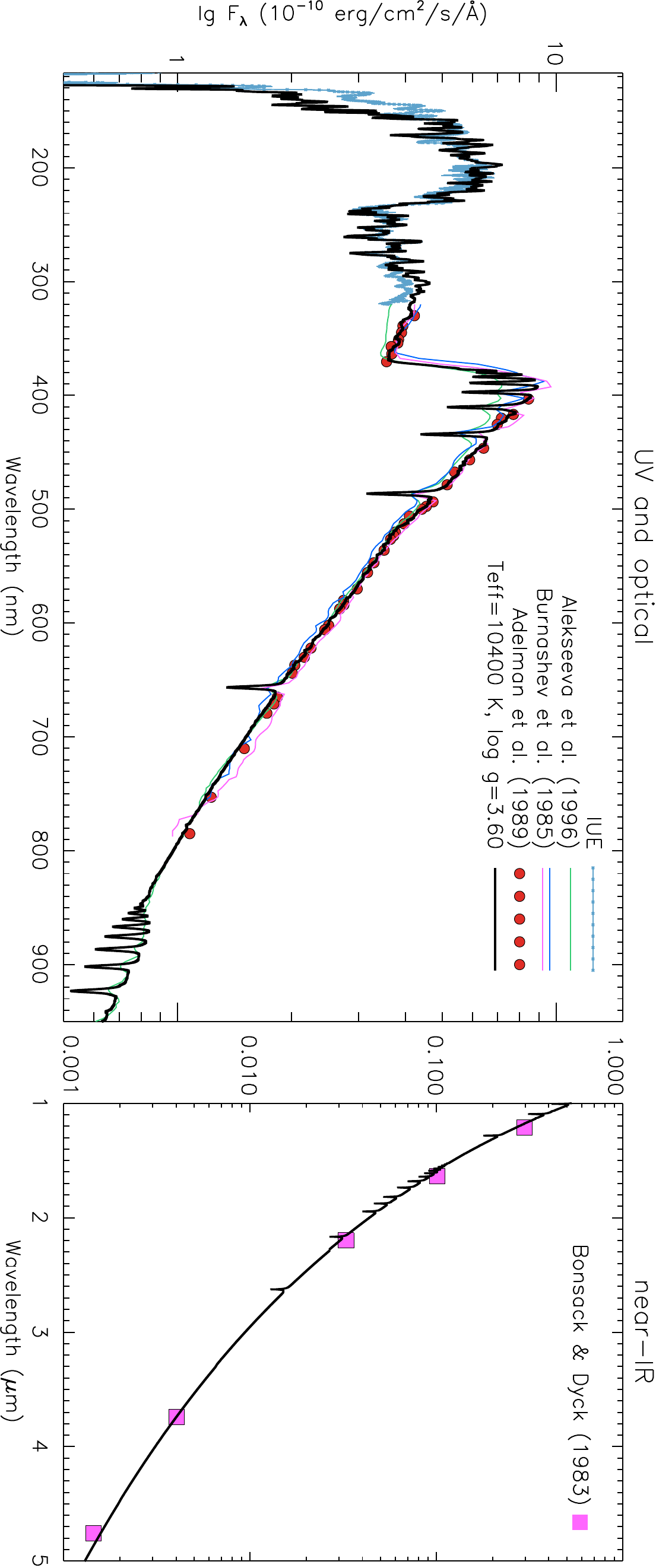}
\caption{Comparison of the observed and computed spectral energy distributions of \aur\ in the UV and optical (\textit{left panel}) and near-IR (\textit{right panel}). The sources of the observed spectrophotometry are indicated in the legends. The thick black line shows the best-fitting theoretical SED corresponding to \teff\,=\,10400~K, \lgg\,=\,3.6, and $\theta$\,=\,0.85~mas.}
\label{fig:sed_aur}
\end{figure*}

\section{Stellar parameters}
\label{params}

Here we revise atmospheric and fundamental parameters of both targets using our high-resolution phase-averaged Narval spectra and spectral energy distributions (SED) covering a wide wavelength range from the UV to the near-IR taken from the literature. Besides improving \teff\ and \lgg\ estimates, this analysis provides a tight constraint on the inclination angle of the stellar rotational axis, $i$, required for surface mapping.

We started our analysis of \aur\ by adopting the stellar parameters (\teff\,=\,10500~K, \lgg\,=\,3.6, \vs\,=\,55~\kms) and mean abundances from the study by \citet{krticka:2015}. The corresponding model atmosphere was computed with the help of the {\sc LLmodels} code \citep{shulyak:2004}. We then employed this model to derive mean abundances by fitting {\sc Synth3} \citep{kochukhov:2007d} synthetic spectra to the phase-averaged Stokes $I$ Narval observations. Depending on the spectral line, visual or automatic, least-squares fitting was carried out with the help of the {\sc BinMag} IDL GUI interface\footnote{\url{http://www.astro.uu.se/~oleg/binmag.html}} \citep{kochukhov:2018}. These spectrum synthesis calculations were based on a line list retrieved from the {\sc VALD3} database \citep{ryabchikova:2015}. We were able to estimate abundances of 12 chemical elements (see the second column of Table~\ref{tbl:abund}) using about 30 individual lines and narrow spectral regions.

\begin{figure}[!th]
\centering
\figps{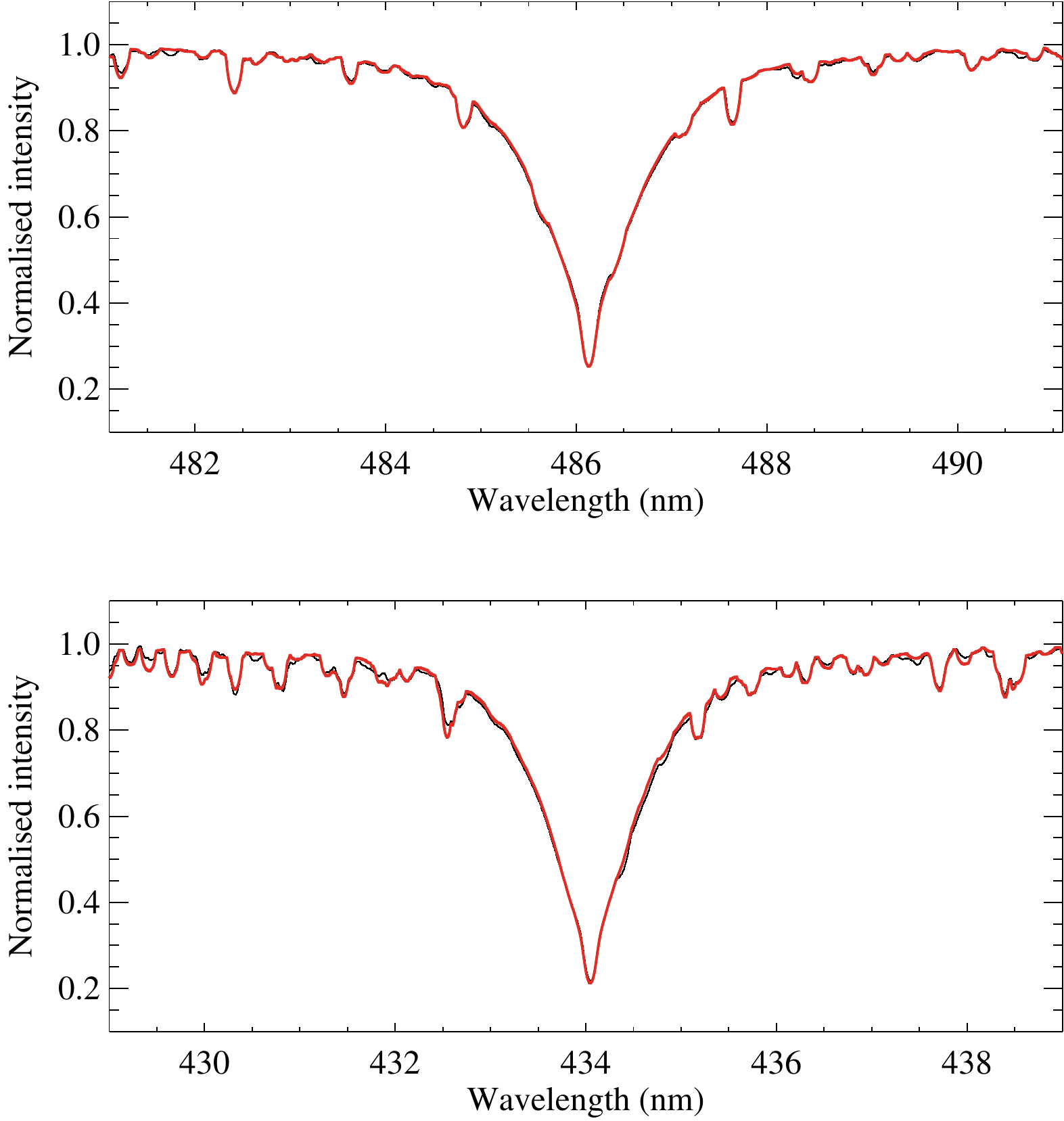}
\caption{Comparison of the average observed (\textit{thin black line}) and computed (\textit{thick red line}) hydrogen H$\beta$ and H$\gamma$ profiles of \aur.}
\label{fig:hlines_aur}
\end{figure}

In the next step we calculated a grid of {\sc LLmodels} atmospheres for different \teff\ and \lgg\ values around \teff\,=\,10500~K, \lgg\,=\,3.6 with the individual element abundances determined above and assuming solar concentrations \citep{asplund:2009} for other elements. Theoretical flux distributions predicted by these models were compared to the composite observed stellar SED obtained by combining the mean {\sc IUE}  {\sc INES} low-resolution, large-aperture spectra \citep{rodriguez-pascual:1999}, the optical spectrophotometry from \citet{burnashev:1985}, \citet{adelman:1989}, \citet{alekseeva:1996} and the near-IR $JHKLM$ photometry by \citet{bonsack:1983} converted to absolute fluxes. This comparison, illustrated in Fig.~\ref{fig:sed_aur}, yields effective temperature \teff\,=\,$10400\pm300$~K and angular diameter $\theta=0.85\pm0.03$~mas. Taking into account the Hipparcos trigonometric parallax $\pi=19.70\pm0.16$ mas \citep{van-leeuwen:2007}, we find $R=4.64\pm0.17$~$R_\odot$. 

The surface gravity of \aur\ was determined by matching the H$\beta$ and H$\gamma$ lines in the average Narval spectrum with the {\sc Synth3} calculations. An excellent fit to both hydrogen lines, shown in Fig.~\ref{fig:hlines_aur}, is obtained with \lgg\,=\,$3.6\pm0.1$. This figure also shows that the intensities of numerous metal lines (mostly Cr and Fe) located in the hydrogen line wings are well-reproduced by the model spectra, confirming our abundance analysis results.

\begin{table}[!t]
%\centering
\caption{Mean element abundances of \aur\ and \uma.
\label{tbl:abund}}
\begin{tabular}{lccr}
\hline
\hline
Element & \aur\ & \uma\ & Sun \\
\hline
He & $-2.3$ & & $-1.11$ \\
C & & $-5.0$ & $-3.61$ \\
O & $-3.5$ & $-3.9$ & $-3.35$ \\
Na & & $-5.4$ & $-5.80$ \\
Mg & $-5.1$ & $-4.6$ & $-4.44$ \\
Si & $-3.2$ & $-5.3$ & $-4.53$ \\
Ca & $-6.5$ & $-6.5$ & $-5.70$ \\
Sc & & $-9.7$ & $-8.89$ \\
Ti & $-7.5$ & $-7.1$ & $-7.09$ \\
Cr & $-4.5$ & $-5.0$ & $-6.40$ \\
Mn & $-5.0$ & $-5.6$ & $-6.61$ \\
Fe & $-3.5$ & $-3.9$ & $-4.54$ \\
Sr & $-8.4$ & $-9.5$ & $-9.17$ \\
Y & & $-9.0$ & $-9.83$ \\
Ba & & $-9.6$ & $-9.86$ \\
Pr & $-8.8$ & $-9.8$ & $-11.32$ \\
Nd & $-7.6$ & $-8.6$ & $-10.62$ \\
\hline
\end{tabular}
\tablefoot{Stellar abundances are given in the $\log (N_{\rm el}/N_{\rm tot})$ units. The corresponding solar abundances are taken from \citet{asplund:2009}.}
\end{table}

\begin{figure*}[!th]
\centering
\fups{16cm}{90}{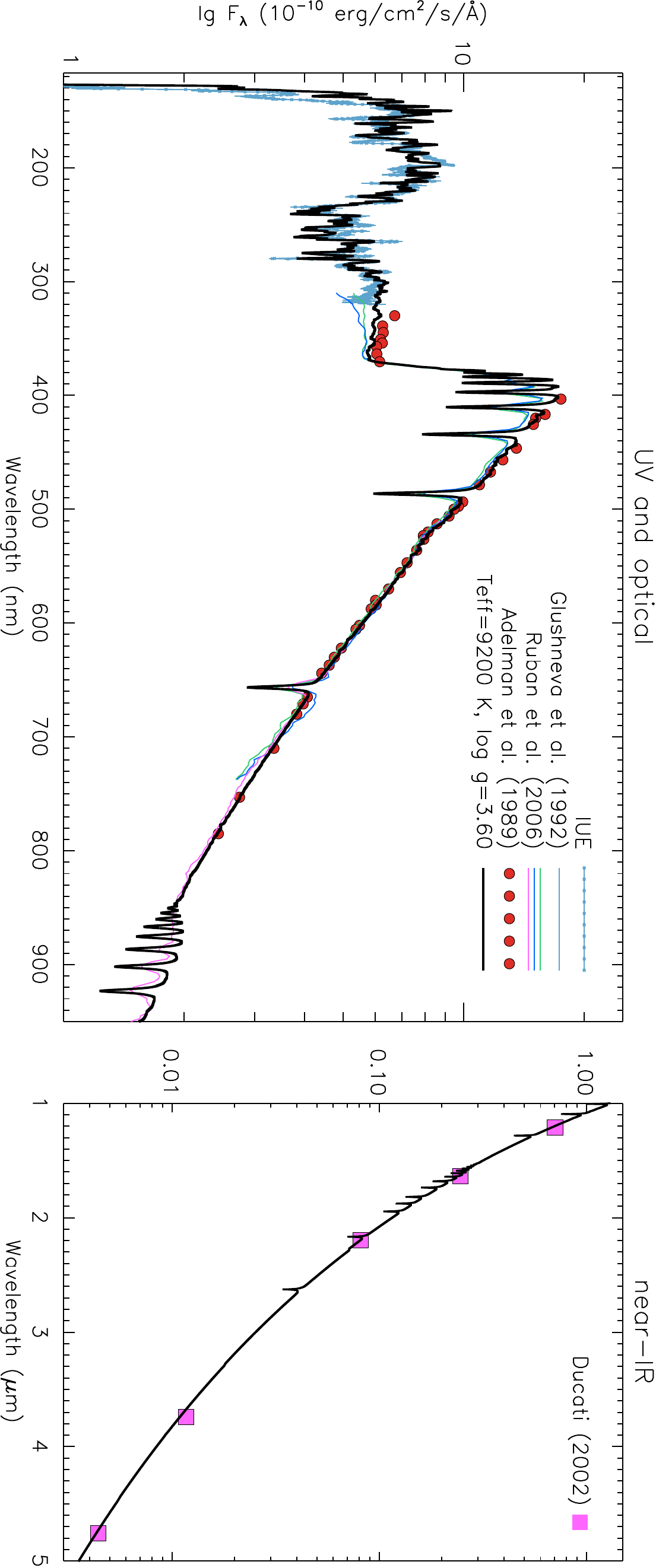}
\caption{Same as Fig.~\ref{fig:sed_aur} for \uma. The best-fitting theoretical SED corresponds to \teff\,=\,9200~K, \lgg\,=\,3.6, and $\theta$\,=\,1.50~mas.}
\label{fig:sed_uma}
\end{figure*}

\begin{figure}[!th]
\centering
\figps{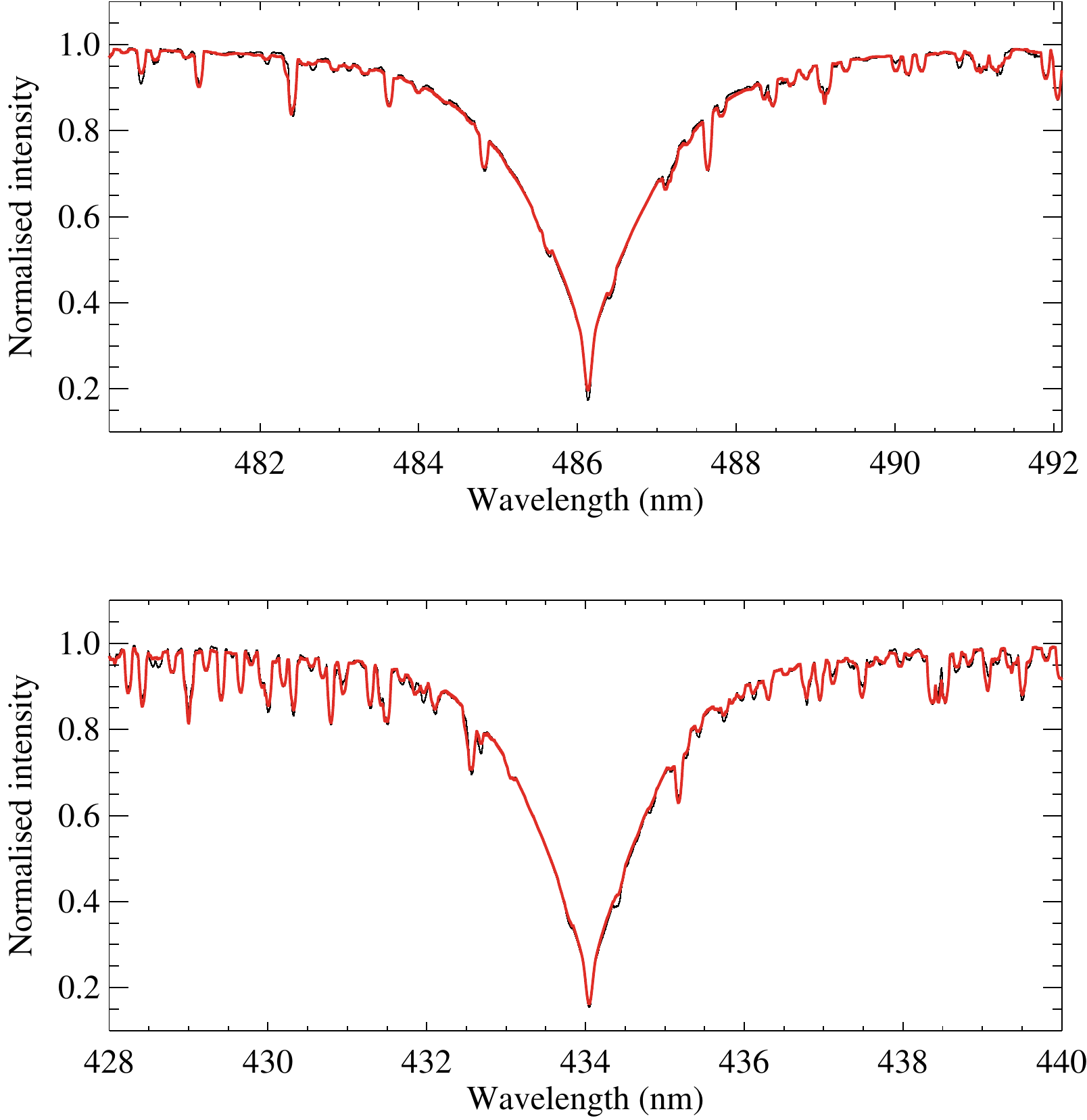}
\caption{Same as Fig.~\ref{fig:hlines_aur} for \uma.}
\label{fig:hlines_uma}
\end{figure}

Using the radius determined above together with the stellar rotational period $P_{\rm rot}=3\fd618664$ and the projected rotational velocity \vs\,=\,$54.0\pm1.0$~\kms\ found in the ZDI analysis below, we infer $i=56\fdg3\pm3\fdg5$. All parameters of \aur\ determined in this study are summarised in the second column of Table~\ref{tbl:params}.

The same parameter determination procedure as described above for \aur\ was applied to \uma. An initial estimate of the mean abundances and stellar parameters (\teff\,=\,9000~K, \lgg\,=\,3.5, \vs\,=\,35~\kms) was adopted from \citet{lueftinger:2003} and \citet{shulyak:2010b}. Owing to its smaller \vs, more lines in the spectrum of \uma\ are suitable for abundance determination. We therefore constrained abundances of 16 elements (see third column of Table~\ref{tbl:abund}) based on the spectrum synthesis modelling of 56 individual lines. The resulting abundance table was employed for calculation of a {\sc LLmodels} atmosphere grid around the initial \teff\ and \lgg. The model fluxes were then fitted to the observed stellar SED, which was constructed by combining the {\sc IUE} data, optical spectrophotometry \citep{adelman:1989,glushneva:1992,ruban:2006}, and near-IR photometry \citep{ducati:2002}. The observed flux distribution is best reproduced by the model with \teff\,=\,$9200\pm200$~K and $\theta=1.50\pm0.05$~mas (Fig.~\ref{fig:sed_uma}). The hydrogen Balmer lines yield \lgg\,=\,$3.6\pm0.1$ (Fig.~\ref{fig:hlines_uma}).

\begin{table}[!th]
%\centering
\caption{Parameters of \aur\ and \uma\ derived in this study.
\label{tbl:params}}
\begin{tabular}{lll}
\hline
\hline
Parameter & \aur\ & \uma\ \\
\hline
\teff\ [K] & $10400\pm300$ & $9200\pm200$ \\
\lgg\ [cgs] & $3.6\pm0.1$ & $3.6\pm0.1$ \\
$\theta$ [mas] & $0.85\pm0.03$ & $1.50\pm0.05$ \\
$R$ [$R_\odot$] & $4.64\pm0.17$ & $4.08\pm0.14$ \\
\vs\ [\kms] & $54.0\pm1.0$ & $35.0\pm0.5$ \\
$i$ [\degr] & $56.3\pm3.5$ & $59.6\pm3.6$ \\
\hline
\end{tabular}
\end{table}

Considering the Hipparcos parallax of \uma, $\pi=39.51\pm0.20$~mas \citep{van-leeuwen:2007}, we determined $R=4.08\pm0.14$~$R_\odot$. This stellar radius, rotational period $P_{\rm rot}=5\fd088631$, and \vs\,=\,$35.0\pm0.5$~\kms\ found below correspond to $i=59\fdg6\pm3\fdg6$. The parameters of \uma\ are summarised in the third column of Table~\ref{tbl:params}.

\section{Longitudinal magnetic field}
\label{bz}

\subsection{Least-squares deconvolved Stokes profiles}

Polarimetric line-addition techniques have proven to be highly effective for boosting the SNR of weak polarisation signatures by taking advantage of redundant line shape information available thanks to the wide wavelength coverage of modern echelle spectropolarimeters. Multi-line methods facilitate measurements of integral magnetic observables, such as the mean longitudinal magnetic field, and provide observational data suitable for detailed line profile modelling with ZDI. Here we apply the technique of least-squares deconvolution \citep[LSD,][]{donati:1997,kochukhov:2010a} to the four Stokes parameter spectra of \aur\ and \uma. 

The LSD method assumes that each line can be represented by a shifted and scaled copy of the mean profile and that spectral contributions of overlapping lines add up linearly. One can invert this simple description, mathematically equivalent to convolution of a line mask and a mean profile in the velocity space, and derive a high-SNR average profile from observations with a series of straightforward matrix operations. In this study we calculate LSD profiles with the help of the {\sc iLSD} code \citep{kochukhov:2010a} and based on the information on the line positions, strengths and polarimetric sensitivities (effective Land\'e factors) extracted from {\sc VALD}. We derived three sets of LSD profiles for each star. The first one was obtained with a line mask including all metal lines which are deeper than 0.1 of the continuum, do not overlap with the hydrogen line wings, and are not affected by telluric absorption. This set of LSD profiles was used for metal line \bz\ measurements with the integral method \citep{wade:2000,kochukhov:2010a}. 

Magnetic CP stars often show significantly different horizontal spot distributions for different chemical elements. These diverse chemical surface structures modulate circular and linear polarisation profiles, which can exhibit substantially different amplitudes and shapes depending on the element considered \citep[e.g.][]{kochukhov:2014,silvester:2014a,yakunin:2015,rusomarov:2018}. For this reason, a detailed, quantitative modelling of magnetic field topologies of MCP stars normally requires using element-specific LSD profiles. With these considerations in mind, we calculated two additional sets of four Stokes parameter LSD profiles for \aur\ and \uma\ based on the line masks containing either Cr or Fe lines (with all other metal lines still taken into account via the second background LSD profile; see \citet{kochukhov:2010a} for details).

The LSD profiles of \aur\ obtained with the full metal line mask containing 2172 individual lines clearly show both circular and linear polarisation signatures. These LSD profiles have a typical uncertainty of $1.2\times10^{-5}$ for Stokes $V$ and $7\times10^{-6}$ for Stokes $QU$, thus yielding a SNR gain of about 50--60 for the same velocity bin compared to the polarisation signatures of individual spectral lines. \bz\ measurements were derived considering the [$-30$, $+93$]~\kms\ velocity interval of the Stokes $I$ and $V$ profiles. The resulting longitudinal field values are reported in Table~\ref{tbl:bz}. The heliocentric Julian dates and rotational phases given in this table correspond to the middle of the circular polarisation observing sequences. \bz\ error bars were inferred from the LSD profile uncertainties, following standard error propagation principles. 

\begin{figure}[!th]
\centering
\fups{8cm}{0}{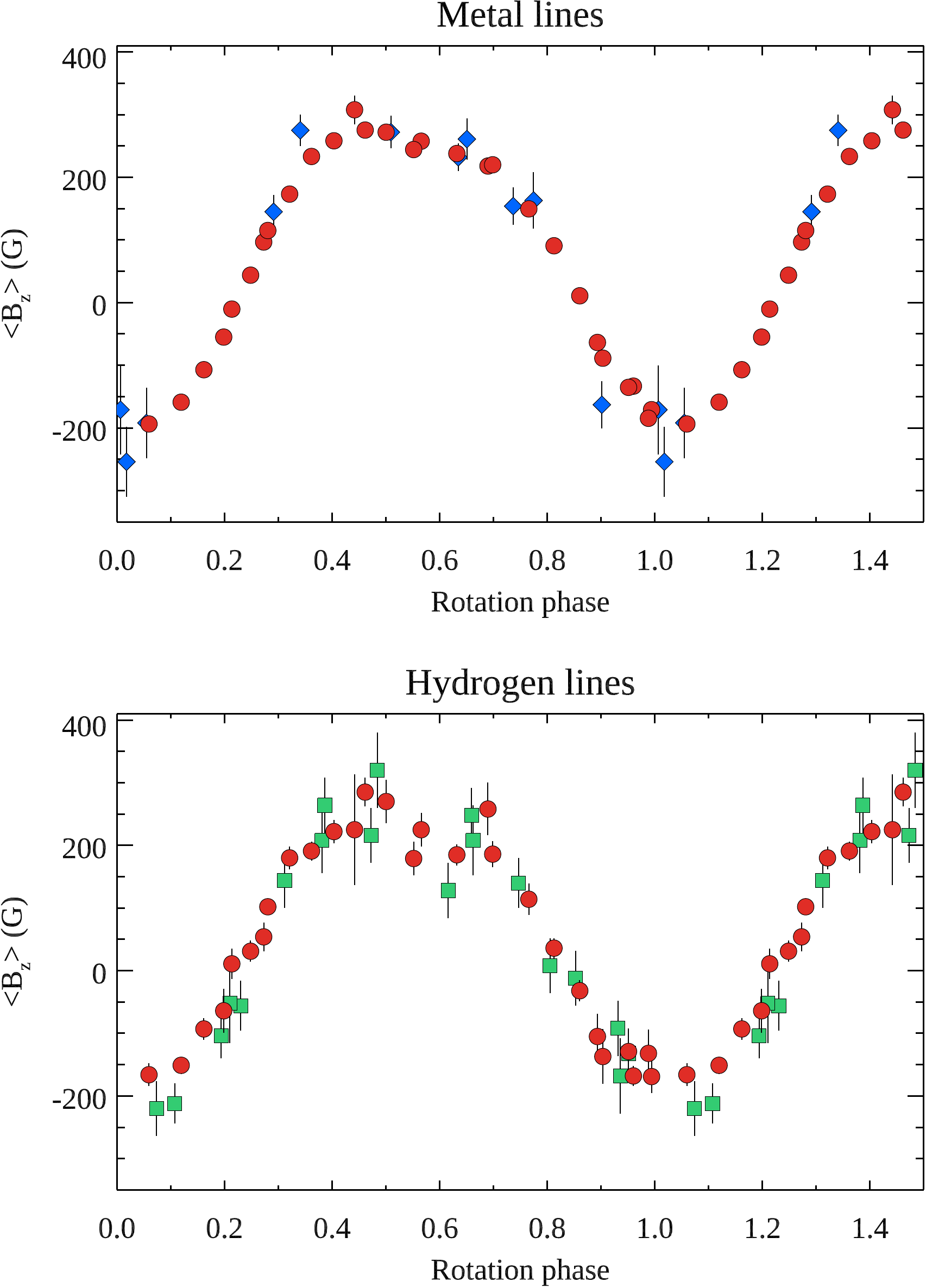}
\caption{Longitudinal magnetic field of \aur\ as a function of rotational phase. The top panel compares the LSD metal line measurements by \citet[][\textit{blue diamonds}]{wade:2000} with the results obtained in our study (\textit{red circles}). The bottom panel compares our Balmer line \bz\ estimates (\textit{red circles}) with the photopolarimetric measurements by \citet[][\textit{green squares}]{borra:1980} corrected by a factor of 4/5.}
\label{fig:bz_aur}
\end{figure}

Our \bz\ measurements range from $-194$~G to $+308$~G and have a median uncertainty of 7~G. These longitudinal field estimates are plotted as a function of rotational phase in the upper panel of Fig.~\ref{fig:bz_aur}. For comparison, previous LSD metal line measurements \citep{wade:2000} are also shown. The resulting \bz\ phase curve has a single-wave character, suggesting that the field topology is dominated by two regions of opposite polarity. However, the curve is also mildly non-sinusoidal, featuring a broad, flattened maximum at phase 0.5 and a narrow minimum at phase 0.0. It is impossible to ascertain whether this distortion is caused by a departure of the stellar magnetic field geometry from a pure dipole or is produced by non-uniform chemical abundance distributions of Fe and Cr, which dominate the LSD line mask.

The Fe LSD profiles of \aur\ were derived from a set of 1354 lines. The mean wavelength and effective Land\'e factor of this line mask, relevant for modelling in Sect.~\ref{zdi}, are $\lambda_0=527.3$~nm and $z_0=1.23$, respectively. The mean Fe Stokes $V$ signatures are detected with high confidence in all observations. These profiles have a typical polarimetric precision of $1.6\times10^{-5}$ and a relative SNR (the peak-to-peak amplitude divided by the mean error) of 31. On the other hand, the Stokes $Q$ signatures are detected with a false alarm probability \citep[FAP,][]{donati:1992} of less than $10^{-3}$ in 18 out of 28 linear polarisation observations. The Stokes $Q$ profiles have a median precision of $9.1\times10^{-6}$ and a SNR of 7. Having a lower amplitude, the Fe LSD Stokes $U$ signatures are detected with FAP\,$<$\,$10^{-3}$ in only 4 observations. 

Another set of element-specific LSD profiles was derived from 417 Cr lines. These profiles, normalised using $\lambda_0=519.9$~nm and $z_0=1.22$, yield somewhat more complex Stokes $V$ signatures, presumably reflecting a more contrasted surface abundance distribution of this element. The Cr Stokes $QU$ signatures have a noticeably higher quality compared to the Fe mean linear polarisation profiles. The Cr $Q$ and $U$ signals are detected in 25 and 20 observations, respectively, with a typical relative SNR of 7--10 in spite of $1.5\times10^{-5}$ polarimetric precision. Both Fe and Cr four Stokes parameter LSD profiles of \aur\ are suited for ZDI inversions.

The metal line LSD Stokes profiles of \uma\ were obtained from a set of 2237 absorption features. The resulting Stokes $V$ spectra have a typical precision of $1.6\times10^{-5}$, corresponding to a factor of 60 gain in SNR. Variable circular polarisation signatures are detected at the FAP level of $<$\,$10^{-5}$ for all but one observation. The typical relative SNR of the Stokes $V$ profiles is 19. On the other hand, no $Q$ or $U$ signatures were detected in any of the observations despite reaching a polarimetric precision of $9\times10^{-6}$.

The longitudinal magnetic field of \uma\ was calculated from the [$-47$, $+32$]~\kms\ velocity interval of the Stokes $IV$ profiles. These \bz\ measurements, given in Table~\ref{tbl:bz}, indicate variation from $-67$ to $+96$~G. The median error of \bz\ measurements is 5~G. Our metal line longitudinal field curve of \uma\ is presented in the upper panel of Fig.~\ref{fig:bz_uma}, where it is compared with the measurements by \citet{wade:2000}. The rotational modulation of \bz\ indicates a predominantly bipolar field geometry. The \bz\ phase curve also exhibits the same slight asymmetry between the positive and negative extrema as noted above for \aur.

The element-specific LSD profiles were derived from 1212 Fe ($\lambda_0=507.6$~nm, $z_0=1.24$) and 449 Cr lines ($\lambda_0=493.9$~nm, $z_0=1.25$). For both sets of profiles the Stokes $V$ polarisation signal is clearly detected for all but one rotational phases. These mean circular polarisation profiles are characterised by a SNR of 12--14 and a precision of 2.2--3.5\,$\times 10^{-5}$. No $QU$ signature detections were achieved with the 1.3--1.9\,$\times 10^{-5}$ noise level. Thus, only the Stokes $IV$ Fe and Cr LSD profile time series can be used for ZDI modelling of \uma. 

\subsection{Hydrogen lines}

\begin{figure}[!t]
\centering
\fups{8cm}{0}{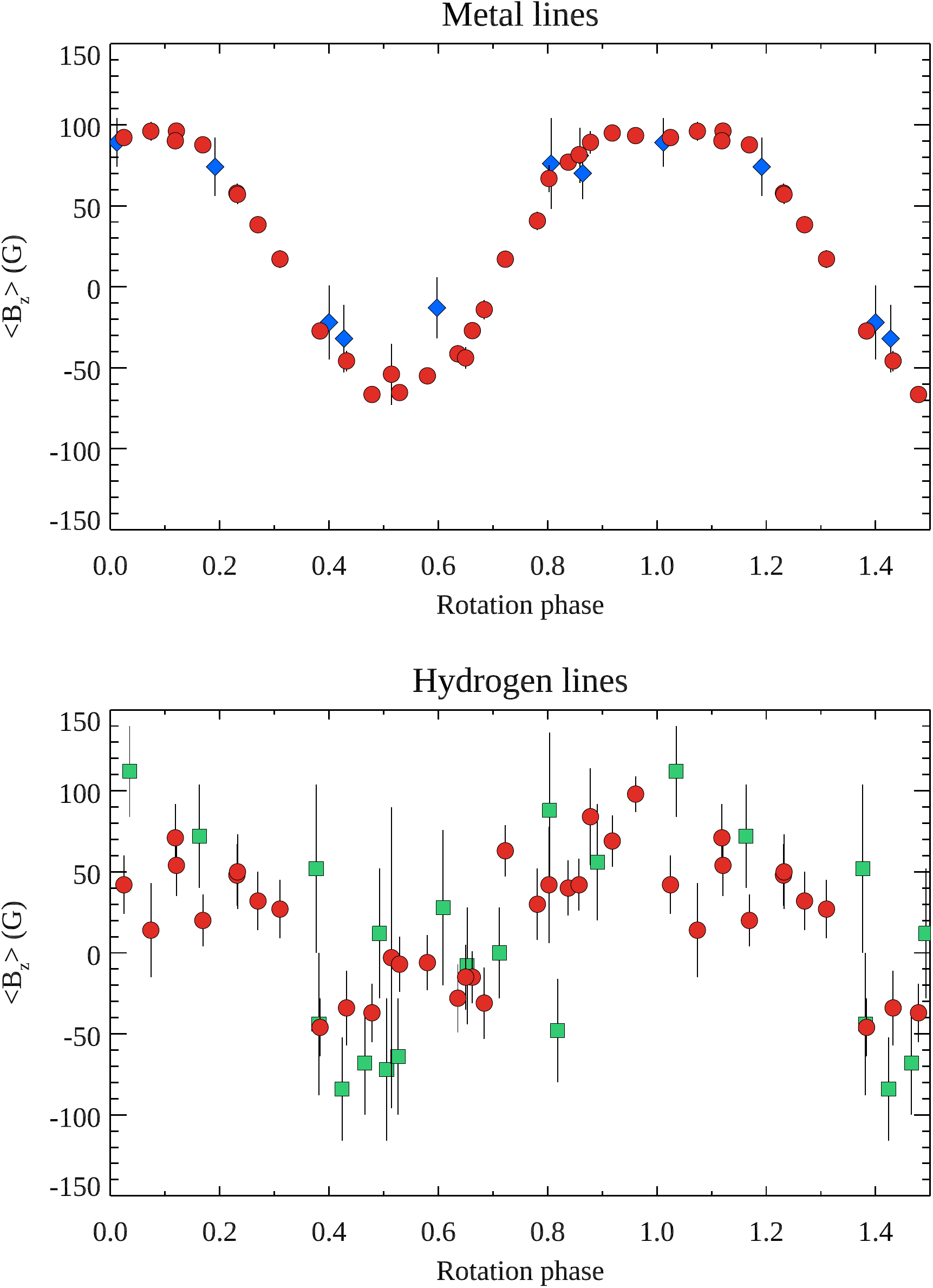}
\caption{Longitudinal magnetic field of \uma\ as a function of rotational phase. The top panel compares our LSD metal line measurements (\textit{circles}) with the measurements by \citet[][\textit{diamonds}]{wade:2000}. The bottom panel shows our Balmer line \bz\ measurements (\textit{circles}) together with the photopolarimetric \bz\ data (\textit{squares}) from \citet{borra:1980} and \citet{bohlender:1990a} corrected by a factor of 4/5.}
\label{fig:bz_uma}
\end{figure}

In addition to the metal line LSD \bz\ measurements of \aur\ and \uma\ presented above, we obtained estimates of the longitudinal magnetic field from the cores of hydrogen Balmer lines \citep{landstreet:2015}. This alternative \bz\ diagnostic is less affected by inhomogeneous surface metal abundance distributions. Results of its application should also be more readily comparable to the historical photopolarimetric \bz\ observations available for both stars \citep{borra:1980,bohlender:1990a}.

\bz\ was measured using the same methodology as described by \citet{shultz:2018a}, i.e. using the centre-of-gravity of Stokes $V$ normalised by the equivalent width (EW) of Stokes $I$, where the EW was measured using the ``line'' continuum at the edges of the rotationally broadened line core rather than the true continuum. In order to improve precision, \bz\ was obtained from the weighted mean of single-line measurements from H$\alpha$, H$\beta$, and H$\gamma$. All three lines return consistent results, with precision declining from H$\alpha$ to H$\gamma$; the median error bar of the weighted mean measurements is about 70\% that of H$\alpha$. 

Our hydrogen core \bz\ measurements of \aur\ are reported in the third column of Table~\ref{tbl:bz} and are shown in the lower panel of Fig.~\ref{fig:bz_aur} together with the photopolarimetric Balmer wing measurements by \citet{borra:1980}, corrected by a factor $4/5$ as recommended by \citet{Mathys:2000} (comparison of historical to modern H line \bz\ measurements of early B-type stars by \citet{shultz:2018a} found that this correction does indeed improve agreement between results obtained using the two different methods). A satisfactory agreement between the new and historical hydrogen \bz\ measurements is evident. The \bz\ curve retains some asymmetry between the positive and negative extrema, indicating the presence of non-dipolar field components.

The hydrogen line \bz\ measurements derived here for \uma\ (Table~\ref{tbl:bz}, lower panel in Fig.~\ref{fig:bz_uma}) have comparable or better precision than the results for \aur\ but appear considerably more noisy owing to the much lower amplitude of the \bz\ curve of the former star. Nevertheless, our results still qualitatively agree with the measurements by \citet{borra:1980} and \citet{bohlender:1990a}. However, not much can be said about detailed shape of the \bz\ phase curve.

\section{Zeeman-Doppler imaging}
\label{zdi}

The Doppler imaging reconstruction of the magnetic field geometries and chemical abundance distributions of \aur\ and \uma\ was carried out using the {\sc InversLSD} magnetic inversion code. This mapping software, developed by \citet{kochukhov:2014} and subsequently used by \citet{rosen:2015}, \citet{kochukhov:2017a}, and \citet{oksala:2018}, is specially designed for accurate, self-consistent modelling of the LSD Stokes parameter profiles of different types of magnetic stars. {\sc InversLSD} allows one to use detailed polarised radiative transfer calculations with realistic model atmospheres and a full line list as an approximation of the local LSD profiles. This approach is more sophisticated and physically sound compared to the single-line Gaussian or Unno-Rachkovsky approximation of the local LSD spectra widely employed by other modern ZDI codes \citep{donati:2006b,morin:2008,folsom:2018}.

A surface mapping calculation with {\sc InversLSD} is split into two main steps. First, we tabulate local theoretical LSD Stokes profiles for a given model atmosphere grid, a range of field strengths, field inclinations with respect to the line of sight, and limb angles. For each node in this grid we compute the full polarised four Stokes parameter spectrum, which covers the entire Narval wavelength range and includes all absorption lines with the intrinsic residual intensity greater than 1\%. We then apply the LSD procedure to these synthetic Stokes spectra, fully consistently (i.e. using the same line mask and observational weights) with the treatment of observations. In the second step, in the course of ZDI least-squares fit of the observed LSD spectra, the local Stokes $IQUV$ LSD profiles are interpolated over the 5-dimensional parameter space (velocity, scalar parameter of the model atmosphere grid, magnetic field strength, field inclination, limb angle), shifted according to the local Doppler velocity and summed taking into account projected surface areas. The $QU$ parameters are additionally transformed according to the local azimuth angle of the field vector. The disk-integrated Stokes parameter profiles are then normalised by the phase-dependent, disk-integrated continuum spectrum. The resulting model spectra are compared to observations at all available rotational phases and the surface distributions are iteratively adjusted to reproduce the data.

In the application of {\sc InversLSD} to MCP stars the scalar parameter of the model atmosphere grid corresponds to the abundance of one particular element (Cr or Fe in this study). This local abundance is implemented both in the polarised spectrum synthesis described above and in the calculations of the underlying {\sc LLmodels} atmospheric grid, allowing us to take into account not only the local equivalent width changes but also the continuum brightness variations as well as atmospheric structure changes associated with chemical spots \citep{kochukhov:2012}.

For the analyses of \aur\ and \uma\ we used local profile grids computed with a 0.25~dex step in the logarithmic Cr and Fe abundance, 25--50~G step in the magnetic field strength and with 15 values in both field inclination and limb angles. The theoretical local LSD profiles were oversampled by a factor of 5 relative to the velocity bin of the observed profiles to ensure an accurate velocity interpolation.

Chemical abundance distributions were parametrised in the usual way, with a discrete two-dimensional longitude-latitude grid containing 1876 surface zones of approximately equal area. Reconstruction of the abundance maps was regularised with the Tikhonov method \citep{piskunov:2002a}, which drives the inverse problem solution to a surface distribution with the least difference between neighbouring surface pixels.

The magnetic field geometry was parametrised using a general spherical harmonic expansion \citep{kochukhov:2014}. In this approach, standard for most recent ZDI studies, the surface vector field distribution is represented as a superposition of the poloidal and toroidal harmonic terms. The free parameters of magnetic mapping problem are the three families of spherical harmonic coefficients, corresponding to the radial poloidal, horizontal poloidal and horizontal toroidal components, with an angular degree running from $\ell=1$ to $\ell=\ell_{\rm max}$ and azimuthal order taking every integer value between $-\ell$ and $+\ell$. We have chosen $\ell_{\rm max}=10$ for both \aur\ and \uma\ to enable reconstruction of complex field structures that can be potentially resolved in the relatively broad line profiles of these stars. This $\ell_{\rm max}$ corresponds to a total of 660 magnetic parameters. However, in practice, we found that the modes with $\ell>6$ for \aur\ and $\ell>3$ for \uma\ contribute less than 1\% of the total magnetic field energy.

The harmonic field model is regularised with a special penalty function \citep{morin:2008,kochukhov:2014}, which favours the low-order modes over the higher-order ones, thus guiding the ZDI solution to the simplest surface field distribution allowed by the data. Both the abundance and magnetic field regularisation parameters need to be adjusted to achieve an appropriate balance between the goals of fitting the data and avoiding the appearance of spurious small-scale surface structures. This adjustment was carried out using the procedure of stepwise regularisation reduction described by \citet{kochukhov:2017}.

\subsection{\aur}
\label{zdi_aur}

The four Stokes parameter ZDI inversions were carried out separately for the Cr and Fe LSD profiles of \aur. We started by determining the best-fitting stellar projected rotational velocity \vs\ and mean radial velocity $V_{\rm r}$ from the Stokes $I$ profiles. This analysis yields \vs\,=\,$54.0\pm1.0$~\kms\ and $V_{\rm r}$\,=\,$31.1\pm0.1$~\kms, respectively. Then we optimised the azimuth angle of the stellar rotational axis $\Theta$, required for modelling linear polarisation observables. This was accomplished by examining the Stokes $QU$ profile fit for different trial values of $\Theta$ ranging from 0\degr\ to 180\degr\ and for the two values of the inclination angle, $i=56\fdg3$ as determined in Sect.~\ref{params} and for the complementary value $i=180\degr - 56\fdg3 = 123\fdg7$. A clear $\chi^2$ minimum was found for $\Theta=80\pm5\degr$ with the latter value of $i$. That $i > 90\degr$ implies that \aur\ is rotating clockwise as seen from the visible (southern) rotational pole.

\begin{figure}[!t]
\centering
\fups{\hsize}{0}{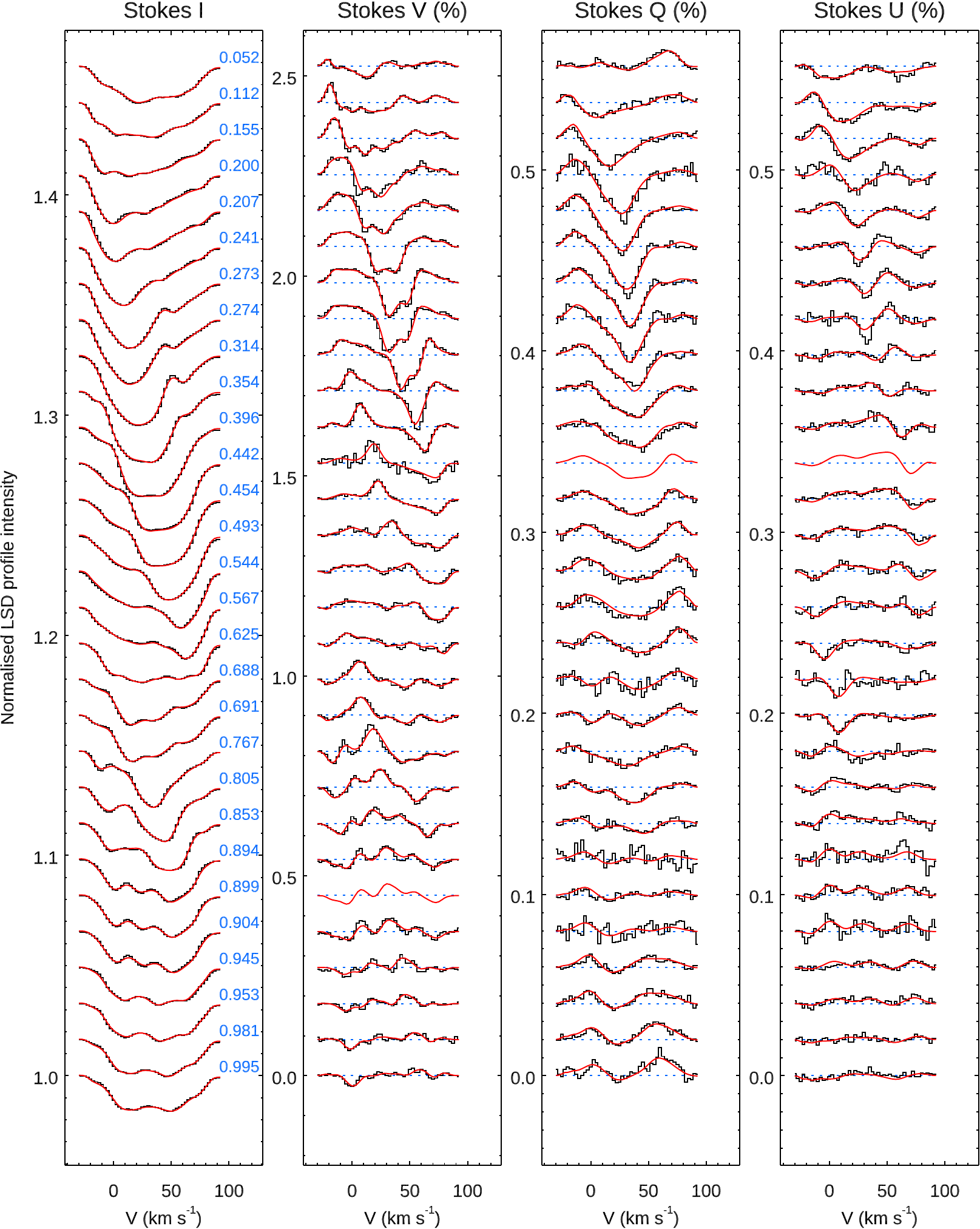}
\caption{Comparison of the observed Cr LSD Stokes $I$, $V$, $Q$, and $U$ profiles of \aur\ with the fit by the magnetic inversion code. Observations are shown with black histograms. Calculations for the final magnetic and chemical spot maps are shown with the solid red lines. Spectra corresponding to different rotation phases are offset vertically. Rotation phases are indicated to the right of each Stokes $I$ spectrum.}
\label{fig:prf_aur_cr}
\end{figure}

\begin{figure}[!t]
\centering
\fups{\hsize}{0}{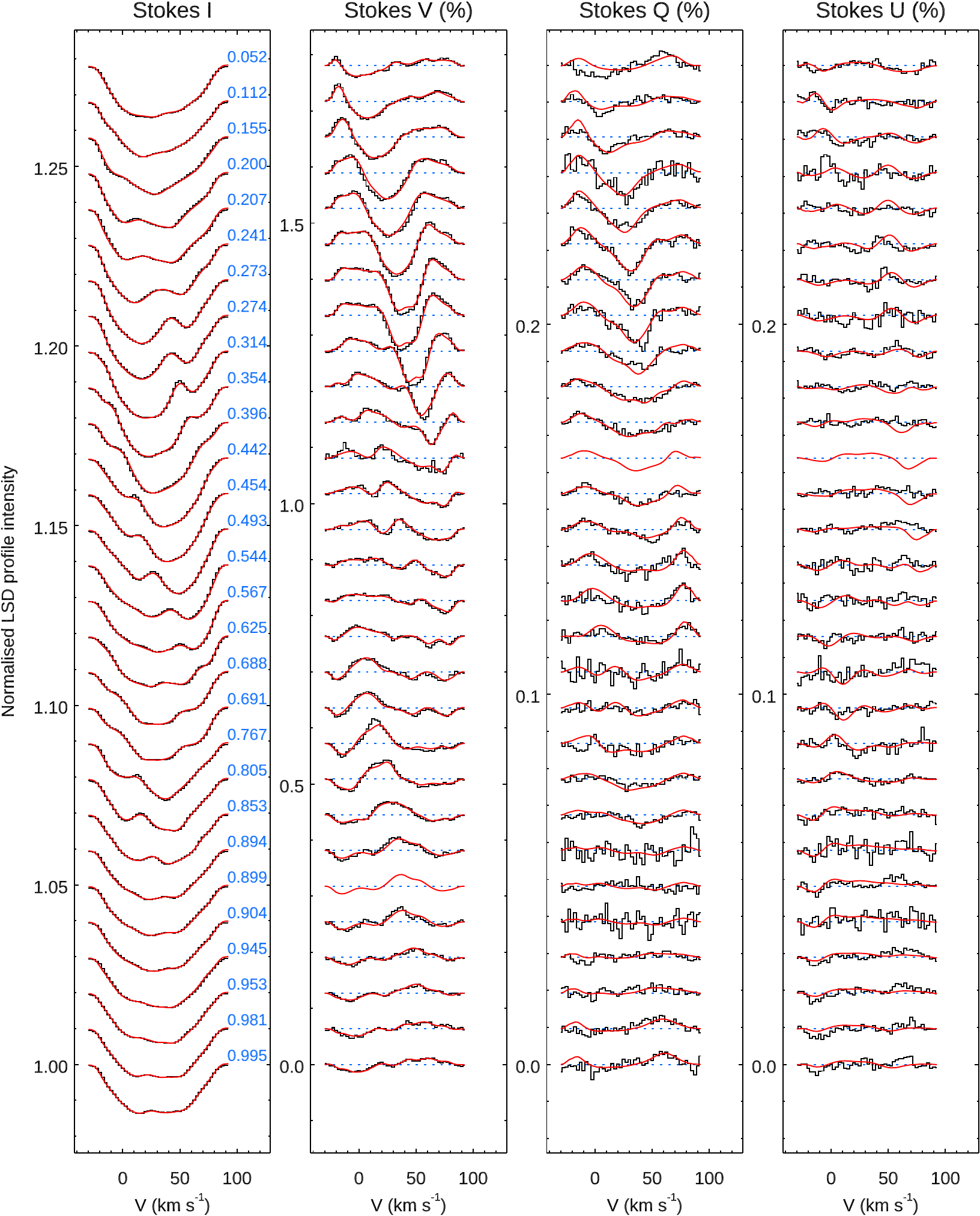}
\caption{Same as Fig.~\ref{fig:prf_aur_cr} for the Fe LSD Stokes $I$, $V$, $Q$, and $U$ profiles of \aur.}
\label{fig:prf_aur_fe}
\end{figure}

\begin{figure*}%[!th]
\centering
\fups{15cm}{-90}{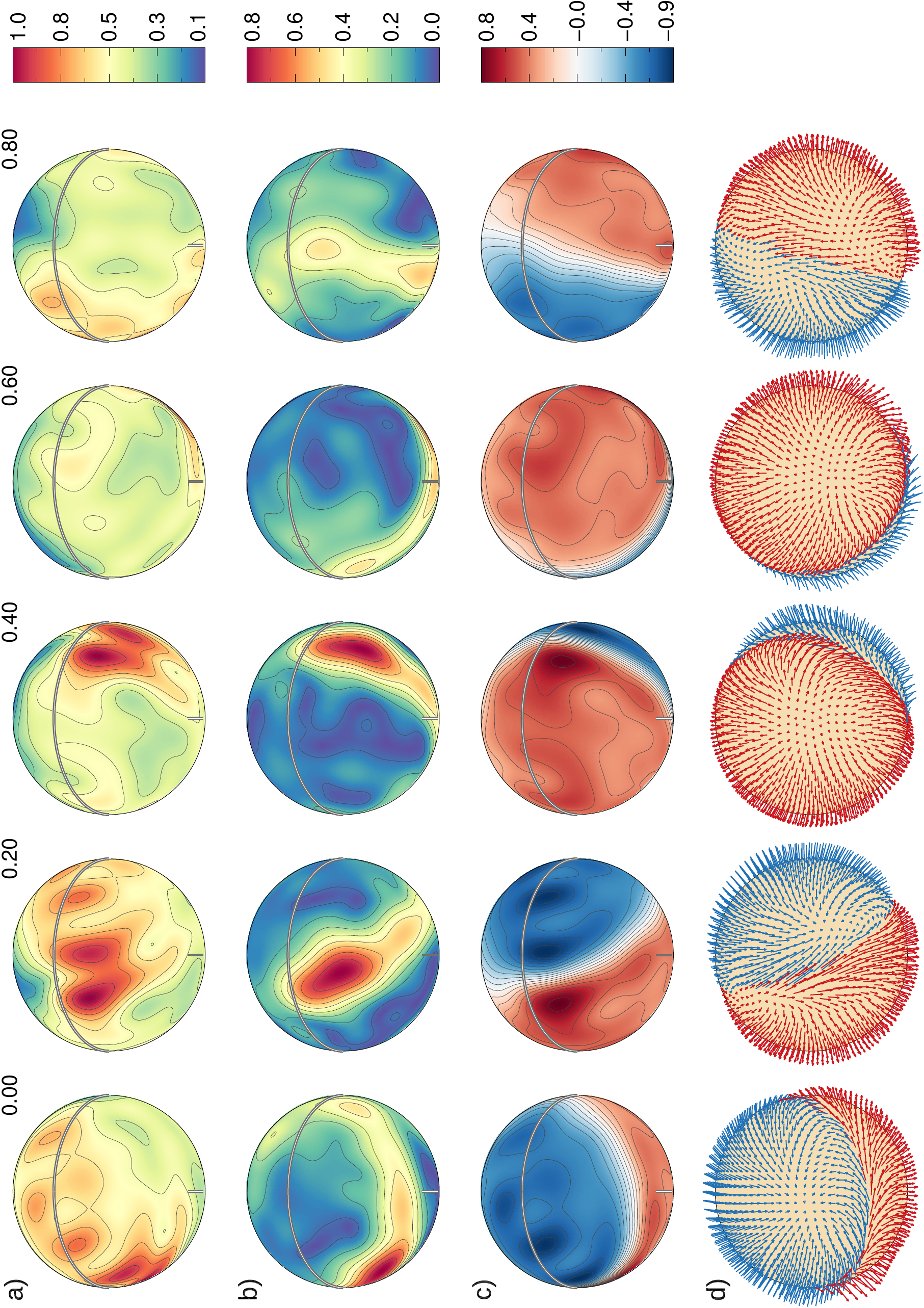}
\caption{Magnetic field topology of \aur\ derived from the Cr Stokes $IQUV$ LSD profiles. The star is shown at five rotation phases, which are indicated above each spherical plot column. The inclination angle is $i=123\fdg7$. The spherical plots show the maps of {\bf a)} field modulus, {\bf b)} horizontal field, {\bf c)} radial field, and {\bf d)} field orientation. The contours over spherical maps are plotted with a step of 0.1~kG. The thick line and the vertical bar indicate the positions of the rotational equator and the visible pole, respectively. The colour bars give the field strength in kG. The two different colours in the field orientation map correspond to the field vectors directed outwards (red) and inwards (blue).}
\label{fig:fld_aur_cr}
\end{figure*}

\begin{figure*}%[!th]
\centering
\fups{15cm}{-90}{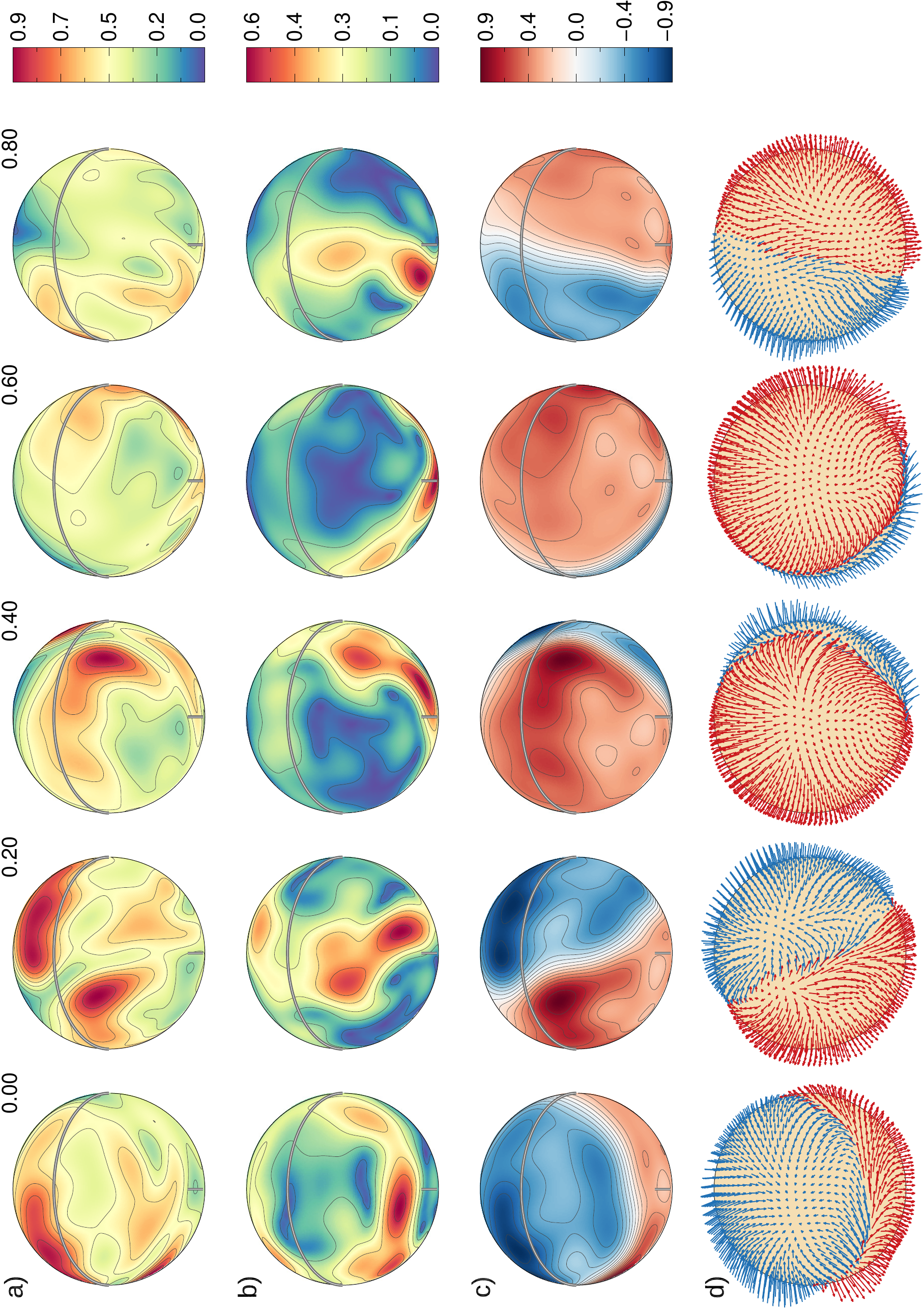}
\caption{Same as Fig.~\ref{fig:fld_aur_cr} for the magnetic field topology of \aur\ derived from the Fe Stokes $IQUV$ LSD profiles.}
\label{fig:fld_aur_fe}
\end{figure*}

The final fit to the observed Stokes $IQUV$ spectra achieved by {\sc InversLSD} is presented in Fig.~\ref{fig:prf_aur_cr} for the Cr LSD profiles and in Fig.~\ref{fig:prf_aur_fe} for Fe profiles. In both cases the observed polarisation data are reproduced within the noise. The corresponding magnetic field maps are illustrated in Figs.~\ref{fig:fld_aur_cr} and \ref{fig:fld_aur_fe}. In these figures the stellar surface is shown in the spherical projection at five different rotational phases and at the actual inclination angle adopted for the inversions. The four rows correspond to the surface maps of the field modulus, horizontal field, radial field, and the field vector orientation. Various statistical characteristics of the Cr and Fe magnetic field maps are reported in Table~\ref{tbl:zdi}.

Our inversion results show that the magnetic field topology of \aur\ has a predominantly dipolar (76--78\% of the magnetic energy concentrated in the $\ell=1$ harmonic components), poloidal (83--91\% energy in the poloidal components) character. The lowest order spherical harmonic component corresponds to a dipolar field strength of 680--700~G. The mean field strength is 440--460~G while the maximum local field strength, including non-dipolar field contributions, is about 0.9--1.0~kG. One can see a number of small-scale deviations from the dipolar geometry, inferred consistently from both Cr and Fe profiles. In particular, the broken ring of stronger horizontal field along the magnetic equator is more pronounced at phase 0.2 than at phase 0.8. The field modulus maps suggest that the strongest magnetic features are, in fact, associated with the magnetic equator rather than the pole. However, the reconstructed morphology of these strongest magnetic spots is somewhat discrepant in the Cr and Fe ZDI maps.

\begin{figure*}%[!th]
\centering
\fups{15cm}{-90}{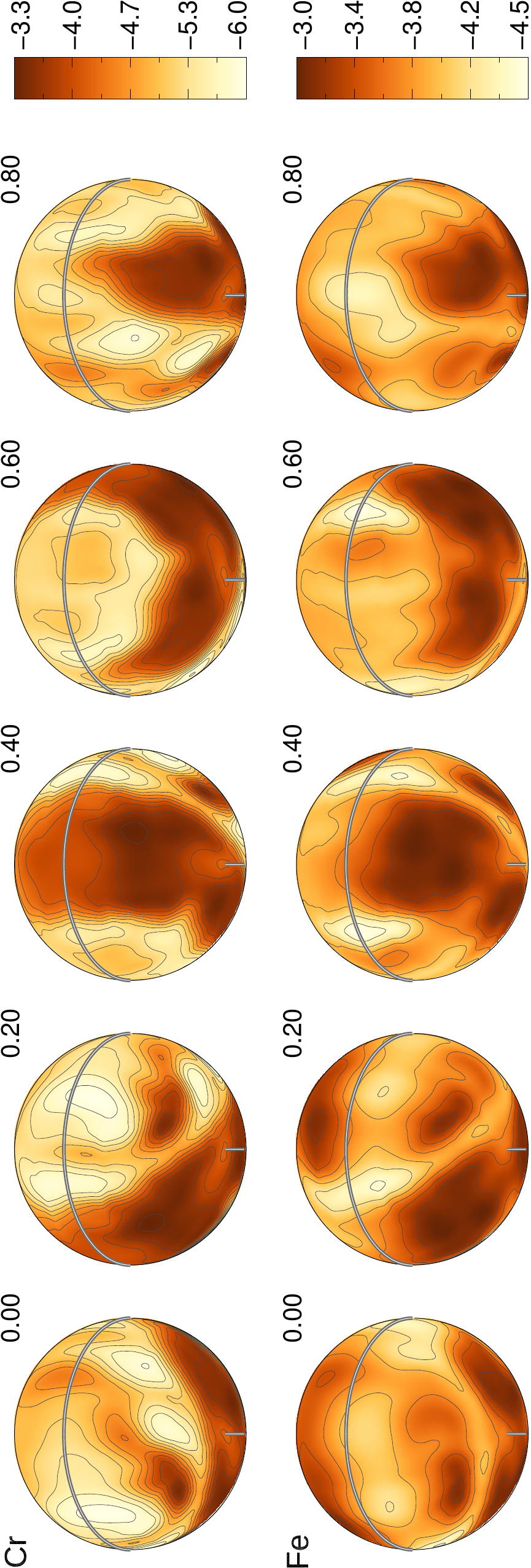}\vspace*{0.2cm}
\fups{14.6cm}{-90}{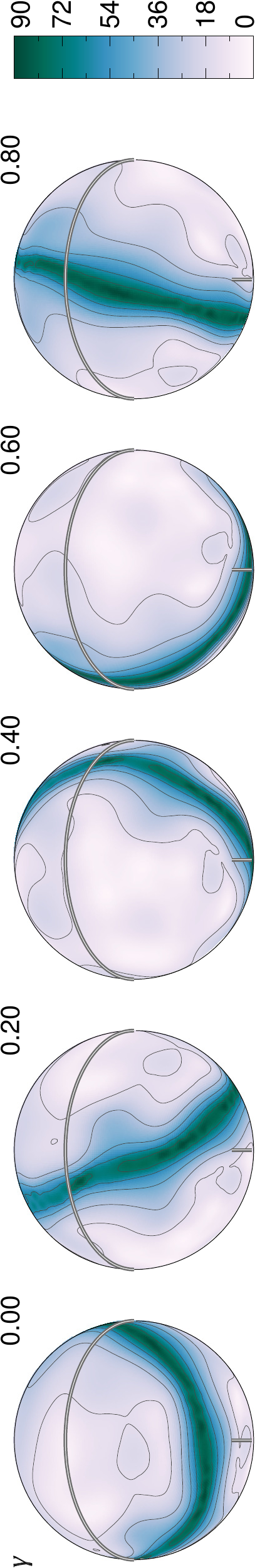}
\caption{Chromium and iron surface abundance distributions (top and middle rows) of \aur\ compared to the local magnetic field inclination (bottom row).
The star is shown at five rotational phases, as indicated next to each plot. The contours are plotted with a 0.2~dex step for the abundance maps and with a 15\degr\ step for the field inclination map. The side colour bars give element abundances in the $\log N_{\rm el}/N_{\rm tot}$ units and field inclination in degrees.}
\label{fig:abn_aur}
\end{figure*}

The inferred deviations from the dipolar field geometry are highly significant. We have verified that the smaller-scale magnetic features are indeed required to properly fit the data by performing test inversions in which the field topology was restricted to a general dipolar field. In the framework of generalised speherical harmonic field parameterisation, $\ell_{\rm max}=1$ still yields 9 magnetic parameters, offering many more degrees of freedom compared to the classical oblique dipole model. However, these inversions still resulted in an inferior fit to the observed polarisation profiles, with the standard deviation increasing by a factor of 2.6--3.1 for Stokes $V$ and 1.7--1.9 for Stokes $QU$ relative to the $\ell_{\rm max}=10$ results reported above.

One can get an idea of the uncertainty of magnetic field reconstruction by examining the difference between the field maps recovered from the Cr and Fe LSD profiles. Restricting this comparison to the more visible latitudes below the rotational equator, we obtained the mean absolute difference of 102, 78, and 94~G for the field modulus, horizontal field, and radial field components, respectively. Thus, the typical uncertainty of our magnetic mapping is on the order of 10\% of the maximum local field intensity. The mean discrepancy of the local field inclination is about 8\fdg5.

The Cr and Fe surface abundance distributions, recovered simultaneously and self-consistently with the corresponding magnetic field geometry maps, are shown in Fig.~\ref{fig:abn_aur}. The bottom row of this figure also illustrates the local field vector inclination (averaged over the Cr and Fe maps) with respect to the surface normal, $\gamma$. This quantity, reckoned in degrees from 0\degr\ (vertical field) to 90\degr\ (horizontal field), is computed from the field modulus $B$ and radial field component $B_{\rm r}$ as $\gamma = \arccos{(|B_{\rm r}|/B)}$. The $\gamma$ map essentially traces horizontal magnetic field regions where theoretical atomic diffusion studies expect the largest accumulation of chemical elements. 

As evident from Fig.~\ref{fig:abn_aur}, both Cr and Fe exhibit highly non-uniform distributions over the surface of \aur, with the local abundance ranging from approximately solar to 1.5--3.0~dex overabundance. The Cr and Fe abundance maps are morphologically similar (though not identical in details), but the contrast is significantly higher for the former element. For both elements the magnetic equator corresponds to narrow rings/arcs of relative underabundance. Areas of higher element concentration are found on both sides of the magnetic equator. The zones located in the middle of the unipolar positive (phase 0.6) and negative (phase 0.0) radial field regions, loosely corresponding to the poles of a dipolar geometry, appear to have a lower element abundance. On average, the negative radial field regions exhibit 0.4--1.0~dex lower abundance of Cr and Fe relative to the zones with positive radial field orientation.

\begin{table}[!t]
%\centering
\caption{Characteristics of the magnetic field topologies of \aur\ and \uma\ derived with ZDI.
\label{tbl:zdi}}
\begin{tabular}{lllll}
\hline
\hline
ZDI map & $B_{\rm mean}$ (G) & $E_{\rm pol}$ (\%) & $E_{\ell=1}$ (\%) & $B_{\rm d}$ (G) \\
\hline
\aur, Cr & 463 & 82.8 & 77.7 & 702 \\
\aur, Fe & 439 & 91.1 & 75.9 & 681 \\
\hline
\uma, Cr & 100 & 95.0 & 85.0 & 163 \\
\uma, Fe & 99 & 95.0 & 87.7 & 170 \\
\hline
\end{tabular}
\end{table}

\subsection{\uma}

Since no usable polarisation signatures were detected in the LSD Stokes $QU$ spectra of \uma, ZDI inversions had to be carried out using only the Stokes $I$ and $V$ observations of that star. In this case, and also owing to a smaller projected rotational velocity of \uma, we expect to reach a somewhat lower spatial resolution of the surface structure details compared to the study of \aur. Nevertheless, the available circular polarisation data with a dense phase coverage is sufficient for establishing main characteristics of the global surface magnetic field and probing its relation to the chemical inhomogeneities.

The chemical element distributions and vector magnetic field maps were derived separately from the sets of Cr and Fe LSD profiles. The best description of the Stokes $I$ line shapes was achieved with \vs\,=\,$35.0\pm0.5$~\kms\ and $V_{\rm r}$\,=\,$-9.1\pm0.1$~\kms. The inclination angle $i=59\fdg6$ was adopted for all inversions according to the results of Sect.~\ref{params}. Due to the lack of linear polarisation constraints, the azimuth angle $\Theta$ cannot be determined and the rotational axis geometries corresponding to $i$ and $180\degr - i$ cannot be distinguished. We therefore assumed $i < 90\degr$, meaning that the star rotates counterclockwise as seen from the visible (northern) rotational pole.

\begin{figure}[!t]
\centering
\fups{4.4cm}{0}{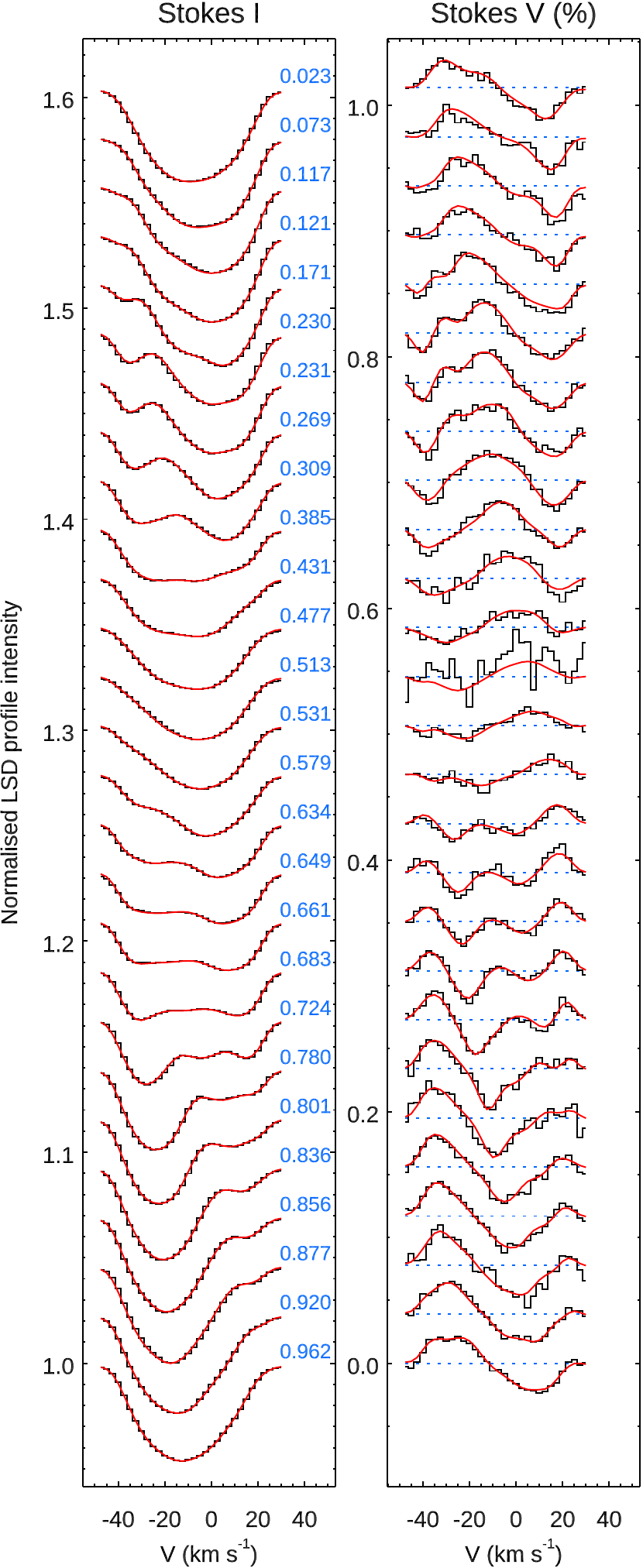}
\fups{4.4cm}{0}{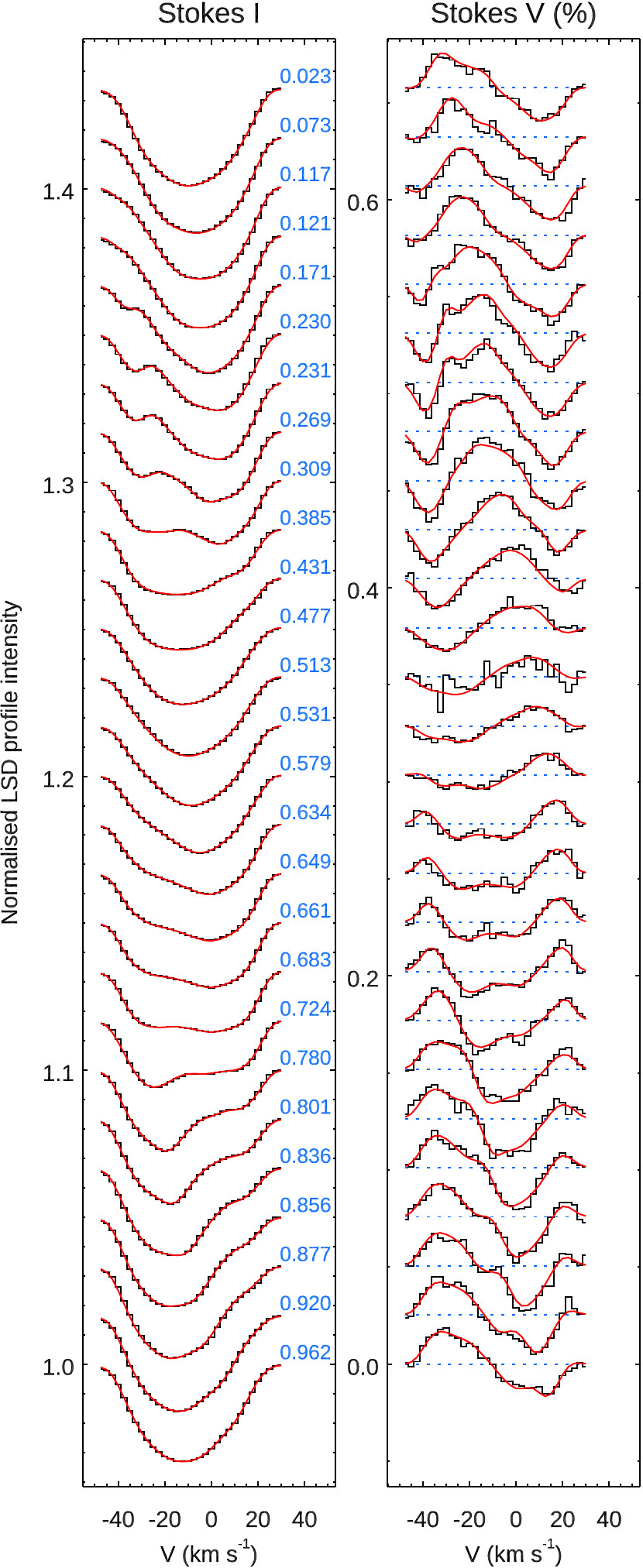}
\caption{Same as Fig.~\ref{fig:prf_aur_cr} for the observed and computed LSD Stokes $I$ and $V$ profiles of \uma. The left pair of panels shows the Cr LSD profiles; the right pair corresponds to the Fe LSD profiles.}
\label{fig:prf_uma}
\end{figure}

\begin{figure*}[!th]
\centering
\fups{15cm}{-90}{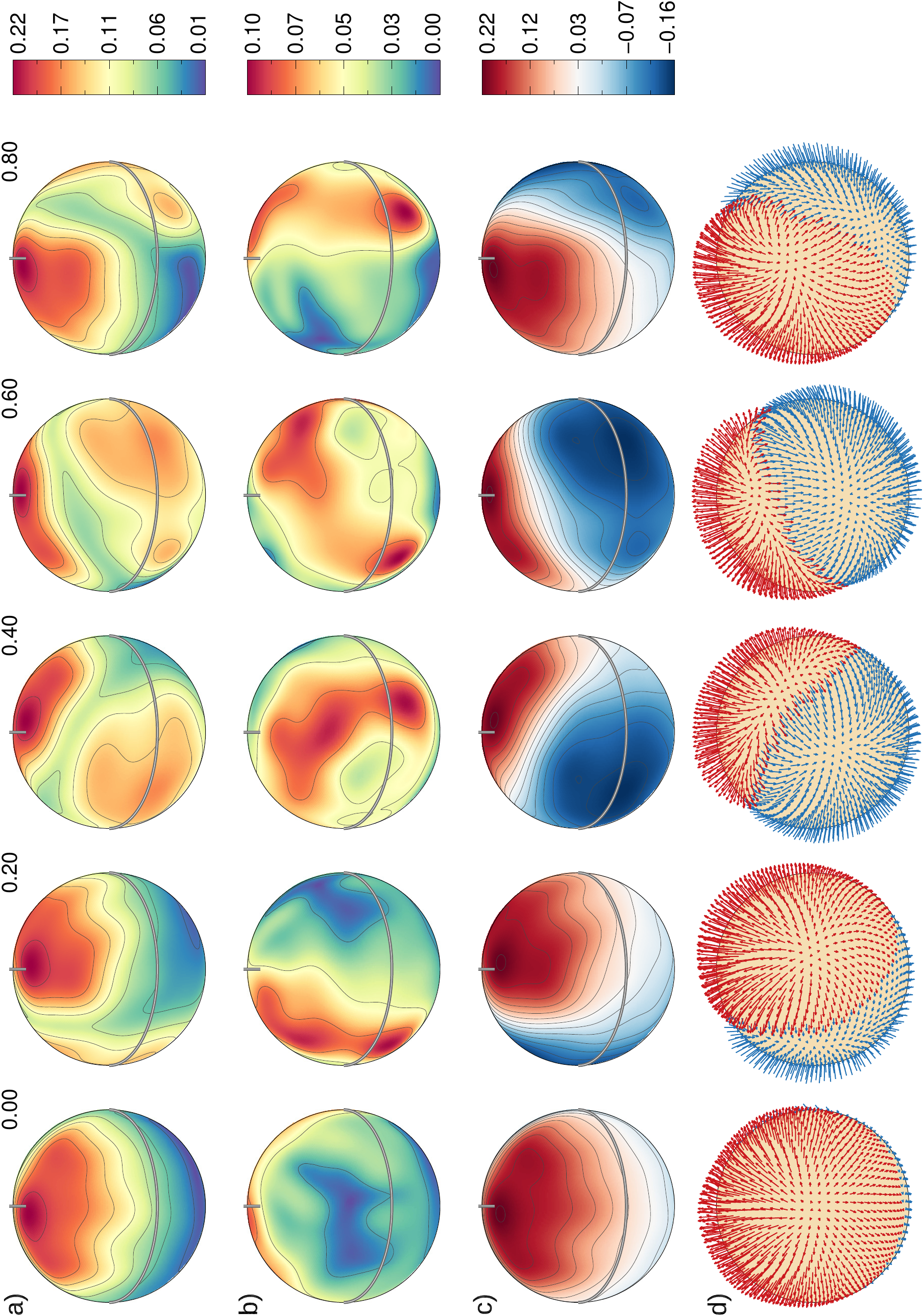}
\caption{Same as Fig.~\ref{fig:fld_aur_cr} for the magnetic field topology of \uma\ derived from the Cr Stokes $IV$ LSD profiles. In this case the contours over spherical maps are plotted with a 0.03~kG step.}
\label{fig:fld_uma_cr}
\end{figure*}

\begin{figure*}[!th]
\centering
\fups{15cm}{-90}{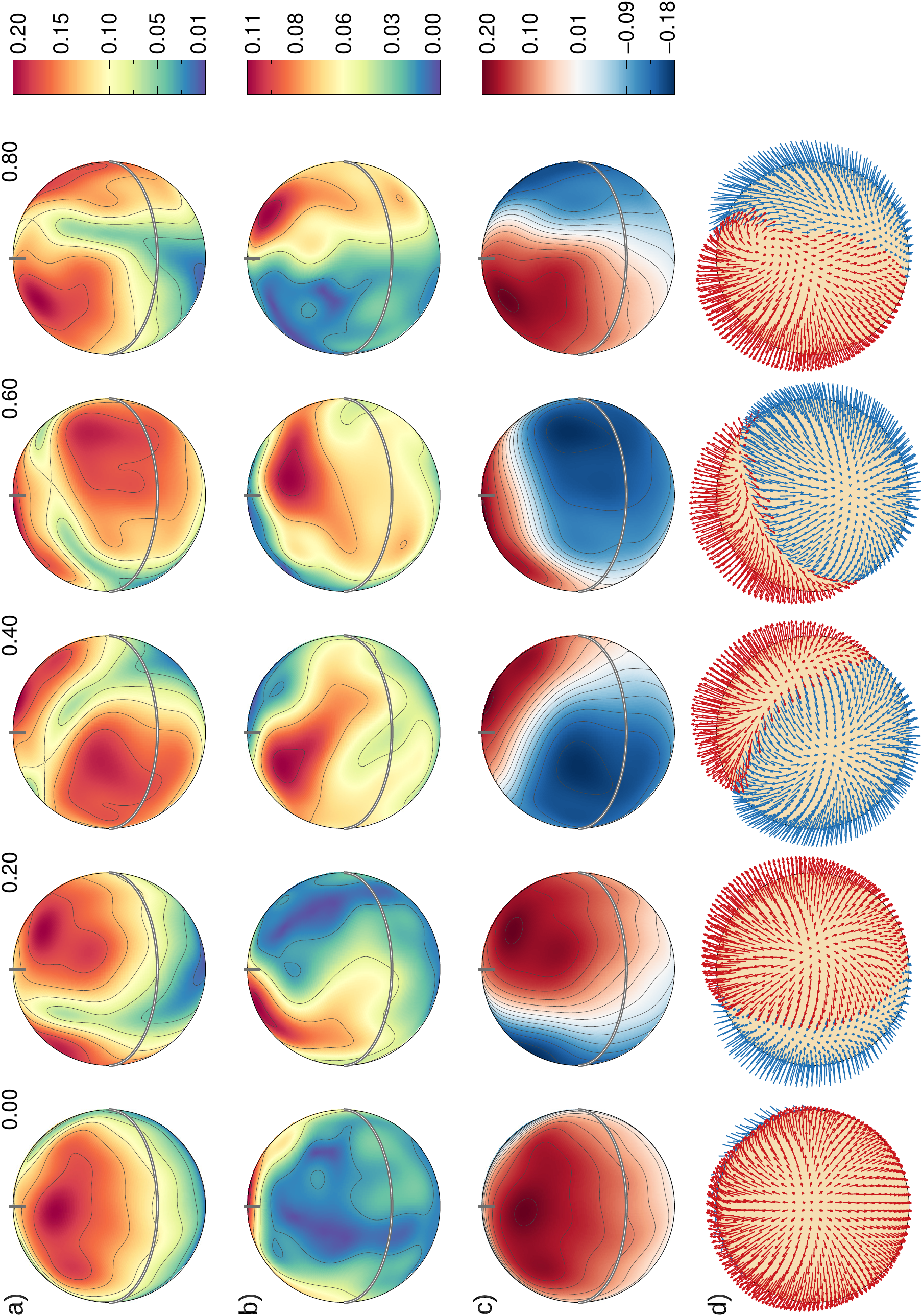}
\caption{Same as Fig.~\ref{fig:fld_uma_cr} for the magnetic field topology of \uma\ derived from the Fe Stokes $IV$ LSD profiles.}
\label{fig:fld_uma_fe}
\end{figure*}

The final fit to the Cr and Fe Stokes $IV$ profiles of \uma\ is presented in Fig.~\ref{fig:prf_uma}. The intensity and circular polarisation observations are successfully reproduced. The corresponding predicted Stokes $QU$ model spectra have typical peak amplitudes of $\approx$\,$10^{-5}$, which is just below the noise level of the observed linear polarisation profiles. The magnetic field geometries of \uma\ inferred from modelling of the two sets of LSD profiles are displayed in Figs.~\ref{fig:fld_uma_cr} and \ref{fig:fld_uma_fe}. Several characteristics of these field topologies are reported in Table~\ref{tbl:zdi}. The global magnetic field structure of \uma\ is evidently dominated by a dipolar component, which comprises 85--88\% of the total field energy. The field geometry is essentially entirely poloidal. The equivalent dipolar field strength is 160--170~G, while the mean field strength is 100~G.

\begin{figure*}[!th]
\centering
\fups{15cm}{-90}{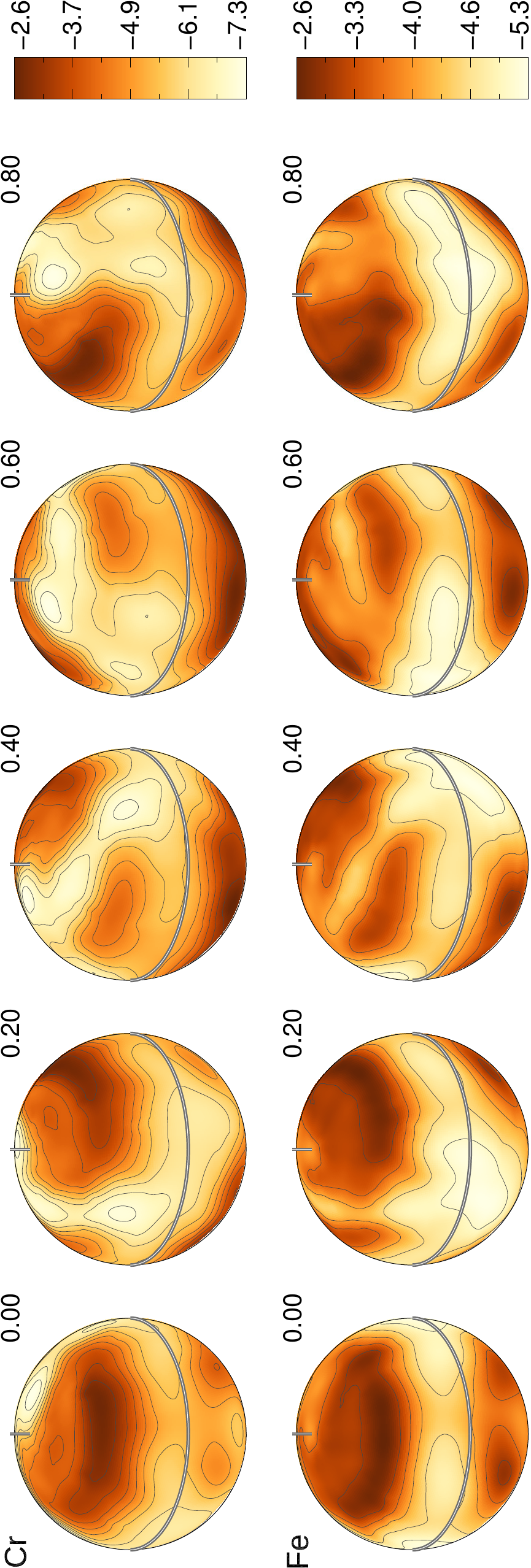}\vspace*{0.2cm}
\fups{14.6cm}{-90}{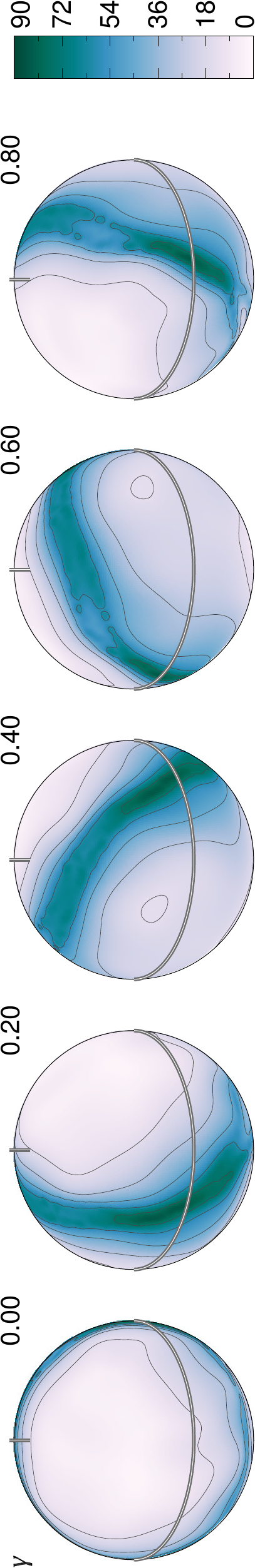}
\caption{Same as Fig.~\ref{fig:abn_aur} for the Cr and Fe surface abundance distributions of \uma\ compared to the local magnetic field inclination.}
\label{fig:abn_uma}
\end{figure*}

The field structure of \uma\ appears to be more nearly dipolar than the global field of \aur. Nevertheless, certain deviations from a purely dipolar geometry are present and are reconstructed consistently from the Cr and Fe profiles. These deviations can be loosely described as an offset of the dipole from the centre of the star. Assuming a general dipolar configuration for the magnetic inversions increases the standard deviation of the Stokes $V$ profile fit by a factor of 1.6, indicating that non-dipolar features of the field topology are statistically significant.

Comparing the two independently reconstructed magnetic field maps, we find mean absolute deviations of 18, 11, and 20~G for the field modulus, horizontal field and radial field components. These values correspond to slightly over 10\% of the peak surface field strength. The average discrepancy of the field vector inclination maps amounts to 13\fdg5.

The Cr and Fe abundance distributions recovered together with the respective magnetic field maps are presented in Fig.~\ref{fig:abn_uma} alongside with the average $\gamma$ map. The inferred local element abundance changes from about 1~dex underabundance relative to the solar chemical composition to an overabundance of approximately 2~dex for Fe and as much as 3.8~dex for Cr. The two surface distributions are very similar, with the Cr map exhibiting a higher contrast compared to Fe. This is expected from the Stokes $I$ profile variability patterns seen in Fig.~\ref{fig:prf_uma}, which are qualitatively similar for the two elements but more pronounced for Cr. Assessing a relation between the chemical spot distributions and the field topology, we find narrow areas of element underabundance at the magnetic equator. There are also relative underabundance zones in the vicinity of the stellar rotational equator, especially for Fe. The most prominent overabundance spots of both elements coincide with the centre of the positive radial field region, best visible at the rotational phase 0.0.

\section{Conclusions and discussion}
\label{conc}

\aur\ and \uma\ are the two brightest and some of the best studied upper main sequence magnetic chemically peculiar stars. In the present study we investigated the magnetic field topologies of these objects using high-quality spectropolarimetric time series observations and an advanced magnetic inversion technique. This analysis provided detailed vector surface magnetic field maps, which represent a key ingredient for modelling the evolution, atmospheres and circumstellar environments of these stars. 

Our ZDI analyses of \aur\ and \uma\ extend simultaneous mapping of the magnetic and chemical abundance surface structures to the previously unexplored regime of weak magnetic field and rapid rotation. In this case, the Zeeman effect produces a negligible impact on the intensity profiles of spectral lines compared to the variation caused by chemical spots. Therefore, abundance DI ignoring the magnetic field yields reliable results. On the other hand, chemical inhomogeneities affect both the intensity and the Stokes $QUV$ polarisation profiles, making magnetic mapping dependent on the spot reconstruction results.

\begin{table*}[!th]
%\centering
\caption{Fundamental parameters and magnetic field characteristics of MCP stars studied with ZDI.\label{tbl:summary}}
{\small
\begin{tabular}{rlrlrrlllcl}
\hline
\hline
HD & Name & $T_{\rm eff}$ & $\lg L/L_\odot$ & $P_{\rm rot}$ & $\langle B \rangle$ & $E_{\rm pol}$ & $E_{\ell=1}$ & $E_{\ell=2}$ & ZDI & Reference \\
number & & (K) & & (d) & (kG) & (\%) & (\%) & (\%) & inversion & \\
\hline
24712  &  DO Eri &    7250 &  0.89 & 12.458 &   2.79 &  99.0 &  98.0 &  1.4  & $IQUV$ & \citet{rusomarov:2015} \\
32633  & HZ Aur &   12800 &  1.98 &  6.430 &  11.27 &  78.8 &  71.2 &  7.3  & $IQUV$ & \citet{silvester:2015} \\
37479  & $\sigma$ Ori E &   23000 &  3.6  &  1.191 &   5.33 &  99.5 &  88.4 & 10.5  &  $IV$  & \citet{oksala:2015} \\
37776  & V901 Ori &   22000 &  3.5  &  1.539 &  12.88 &  78.2 &  10.7 &  9.0  &  $IV$  & \citet{kochukhov:2011a} \\
40312  & $\theta$ Aur &   10400 &  2.35 &  3.619 &   0.44 &  88.8 &  78.7 &  6.2  & $IQUV$ & This study \\
62140  & 49 Cam &    7800 &  1.23 &  4.287 &   2.13 &  79.7 &  51.8 & 10.3  & $IQUV$ & \citet{silvester:2017} \\
65339  & 53 Cam &    8400 &  1.40 &  8.027 &  15.48 &  83.7 &  54.1 & 15.6  & $IQUV$ & \citet{kochukhov:2004d} \\
75049  &               &   10250 &  1.65 &  4.048 &  25.64 &  95.8 &  91.7 &  5.2  &  $IV$  & \citet{kochukhov:2015} \\
79158  & 36 Lyn &   13000 &  2.54 &  3.835 &   1.60 &  63.0 &  90.2 &  5.5  &  $IV$  & \citet{oksala:2018} \\
112185 & $\varepsilon$ UMa &    9200 &  2.03 &  5.089 &   0.10 &  95.2 &  87.0 &  9.5  &  $IV$  & This study \\
112413 & $\alpha^2$ CVn &   11600 &  2.00 &  5.469 &   2.11 &  92.2 &  68.0 &  9.5  & $IQUV$ & \citet{silvester:2014a} \\
119419 & V827 Cen &   11150 &  1.62 &  2.601 &  17.21 &  67.6 &  55.6 &  22.5 & $IQUV$ & \citet{rusomarov:2018} \\
124224 & CU Vir &   12750 &  2.00 &  0.521 &   1.14 &  88.0 &  63.9 &  23.9 &  $IV$  & \citet{kochukhov:2014} \\
125248 & CS Vir &    9850 &  1.62 &  9.296 &   4.40 &  71.4 &  63.4 &  17.7 & $IQUV$ & \citet{rusomarov:2016} \\
133880 & HR Lup &   12000 &  2.10 &  0.877 &   4.01 &  95.3 &  69.7 &  21.8 &  $IV$  & \citet{kochukhov:2017a} \\
149438 & $\tau$ Sco &   32000 &  4.5  & 41.033 &   0.26 &  45.6 &  12.1 &  36.1 &  $IV$  & \citet{kochukhov:2016a} \\
184927 & V1671 Cyg &   22000 &  3.6  &  9.531 &   4.72 &  79.0 &  42.6 &  57.4 &  $IV$  & \citet{yakunin:2015} \\
\hline
\end{tabular}
}
\end{table*}

Compared to the analysis of \uma, the ZDI of \aur\ benefited from the inclusion of Stokes $Q$ and $U$ spectra in the magnetic inversions. Moreover, the latter star has a larger \vs, leading to a higher spatial resolution of tomographic maps. We assessed the impact of both of these effects with simulations described in Appendix~\ref{zdi_test}. Results of this analysis indicate that, in this particular case, neither the \vs\ difference nor availability of the Stokes $QU$ spectra has a major influence on the reconstruction of magnetic field and abundance maps. 

The magnetic field maps derived in this paper indicate that \aur\ and \uma\ have mean field strength of $\approx$\,450 and 100~G, respectively, which is significantly weaker than typically found for nearby MCP stars \citep{power:2008,sikora:2018}. The global magnetic field topology is predominantly dipolar for both objects. Nevertheless, our high-resolution polarisation spectra of these stars cannot be reproduced in detail without allowing for some small-scale deviations from the dipolar geometries. Our assessment shows that these deviations are statistically significant. The distorted dipolar field topologies of \aur\ and \uma\ are qualitatively similar to the surface field structures found by ZDI studies of MCP stars with stronger fields \citep{kochukhov:2014,kochukhov:2015,kochukhov:2017a,oksala:2015,silvester:2015,silvester:2017}. This result suggests that there is no obvious trend of the degree of field complexity with its mean intensity.

Both \aur\ and \uma\ are known to be evolved Ap stars, located near the terminal-age main sequence in the HR-diagram \citep{kochukhov:2006}. This is supported by the large radii and low \lgg\ values inferred in our study. The absence of a dramatic structural difference between the surface fields of younger MCP stars and the two evolved objects investigated here indicates that the field complexity does not strongly depend on age, at least in the 2.8--3.4~$M_\odot$ mass range to which these two stars belong. On the other hand, the large radii of these stars may explain their lower than average magnetic field strengths. Considering the fundamental parameters of \aur\ and \uma, we expect these stars to have had at least 2 times smaller radii at the zero-age main sequence. Assuming magnetic flux conservation, one can estimate that the stars started their main sequence evolution with 2.8 and 0.7~kG dipolar fields for \aur\ and \uma, respectively. This makes the initial field of \aur\ very similar to the 2.5~kG average dipolar field of MCP stars in the solar neighbourhood \citep{power:2008,sikora:2018}. The initial field of \uma\ must have been weaker than is observed for a typical MCP star. But it is still well in excess of the empirical $B_{\rm d}=0.3$~kG fossil field threshold established by \citet{auriere:2007}. 

Considering that a significant number of early-type magnetic stars has been analysed in the last decade using ZDI modelling of high-resolution spectropolarimetric observations, it is of interest to assess these results for the presence of systematic trends between different magnetic field characteristics or between the field morphology and stellar parameters. Some initial analysis along these lines was carried out by \citet{kochukhov:2018c}. Here we present an updated summary of ZDI results.

We collected information on one spectroscopically normal magnetic O-type star ($\tau$~Sco) and 16 MCP stars for which detailed magnetic field maps were produced, including the two Ap stars studied here. For each star we recorded the effective temperature, luminosity (either given directly in individual papers or inferred from the quoted radii and temperatures), and rotation period as well as several magnetic field characteristics (the surface-averaged field strength, fraction of magnetic energy contained in the poloidal component, the dipole and quadrupole energy fractions). These data are listed in Table~\ref{tbl:summary}, where references to original ZDI analyses are also given. Whenever studies derived more than one magnetic field map, for example using lines of different chemical elements, we first averaged all available maps and then derived magnetic field parameters.

\begin{figure*}[!th]
\sidecaption
\fups{12cm}{0}{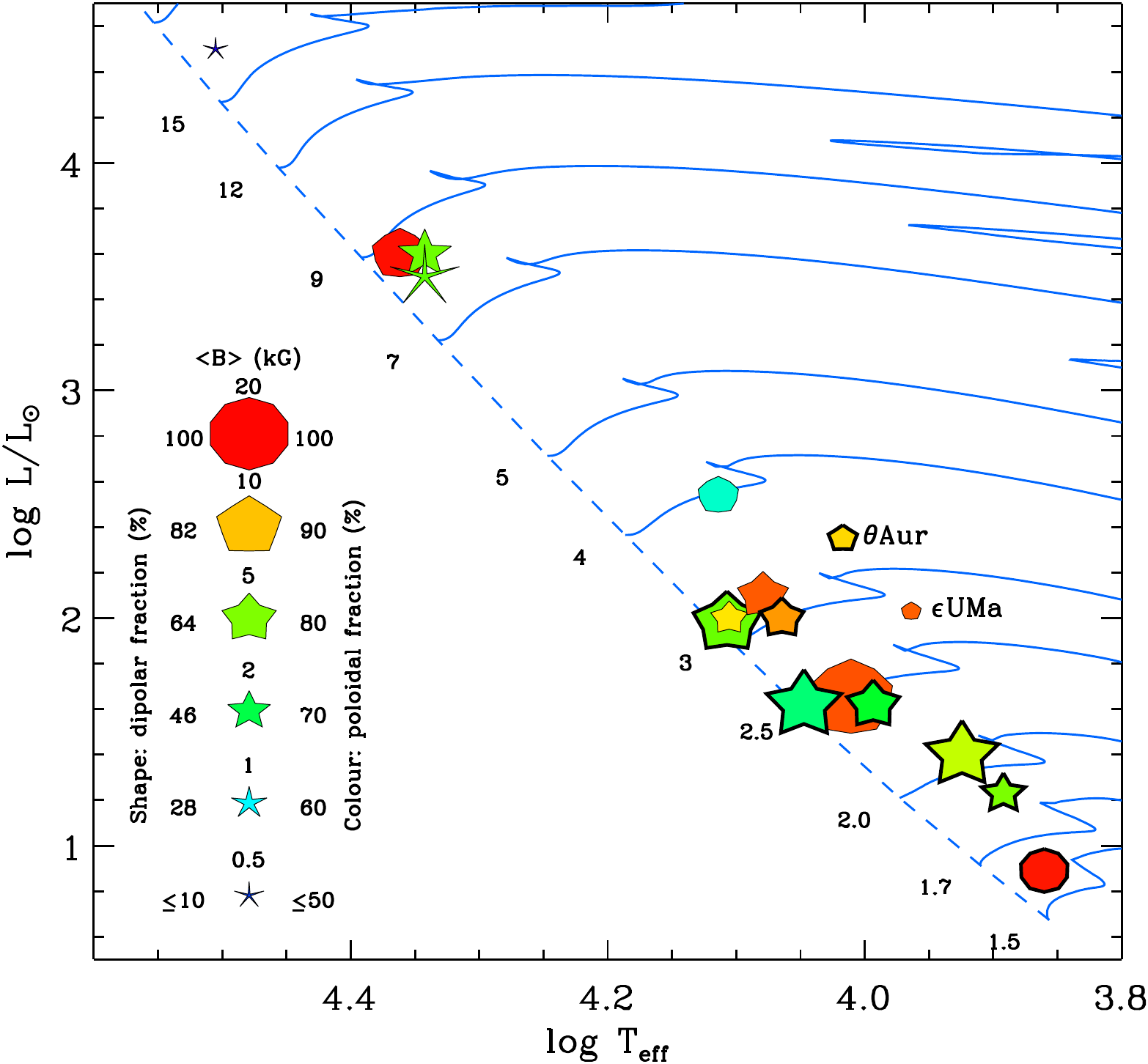}
\caption{Characteristics of the global magnetic field topologies of CP stars studied with ZDI as function of stellar temperature and luminosity. The symbol size indicates the field strength. The symbol shape corresponds to the contribution of the dipolar component to the total magnetic field energy (from decagons for purely dipolar fields to pointed stars for non-dipolar field topologies). The symbol colour reflects the contribution of the toroidal magnetic field component (red for purely poloidal geometries, dark blue for field configurations with $\ge$50\% toroidal field contribution). The thickness of the symbol outline indicates stars studied with the full Stokes vector ZDI (thick outline) or using Stokes $IV$ inversions (thin outline). The theoretical stellar evolutionary tracks shown in the figure \citep{mowlavi:2012} start from the zero-age main sequence (dashed line). The initial stellar masses are indicated next to each track in solar units. The two stars studied in this paper are identified in the plot.}
\label{fig:summary}
\end{figure*}

The graphical summary of ZDI results is shown in Fig.~\ref{fig:summary}. Each MCP star is placed in the H-R diagram, with the symbol size, shape and colour encoding information on the surface magnetic field properties. There are no obvious trends that emerge from this picture. For example, the degree of field complexity does not seem to depend on the stellar mass, with the exception of the fact that conspicuously non-dipolar fields are only found in young massive stars ($\tau$~Sco, HD\,37776). Among the lower mass objects ($M\le4$~$M_\odot$) evolved stars tend to have weaker fields, in accordance with the discussion above. No other dependence on stellar age can be discerned, although it can be argued that analysis of cluster stars (of which there are only a few in our sample) is required to reliably probe evolutionary changes of the global field characteristics \citep{landstreet:2007}. In general, the global geometry of MCP star magnetic fields changes little from one star to another, with nearly all stars showing dominant dipolar fields with varying degree of distortion and addition of smaller scale structures. Detailed line profile analyses fail to confirm the ubiquitous quadrupole-dominated global field topologies inferred by coarse magnetic modelling \citep{landstreet:2000,bagnulo:2002}.

The abundance distributions of Fe and Cr obtained for the two MCP stars studied in our paper exhibit high-contrast patterns, which appear to be more complex than anticipated by atomic diffusion theory. Both equilibrium \citep{leblanc:2009,alecian:2010,alecian:2015} and time-dependent \citep{stift:2016,alecian:2017} theoretical diffusion calculations predict significant vertical abundance gradients in the stellar atmosphere along with a lateral distribution dominated by an accumulation of chemical elements in the horizontal field regions. As demonstrated by \citet{kochukhov:2018a}, the vertical inhomogeneity aspect is largely irrelevant for 2D DI and ZDI studies of fast-rotating MCP stars, implying that, according to the theory, one should find prominent overabundance rings coinciding with magnetic equators for almost all elements in all stars with predominantly dipolar fields. Instead, the Cr and Fe DI maps of both \aur\ and \uma\ exhibit relative depletions of elements at the magnetic equators as well as plenty of other chemical spot structure without an obvious correlation with the local magnetic field. These results contribute to the tension between empirical findings and the current diffusion theory predictions noted by several other recent ZDI studies of MCP stars \citep[e.g.][]{kochukhov:2017a,silvester:2015,silvester:2017}. In the light of this systematic disagreement it would be prudent to revisit the central assumption of atomic diffusion calculations that the local element accumulation is uniquely determined by the local magnetic field parameters. 

The ZDI maps presented here are uniquely suited for testing future improved diffusion computations because the weakness of stellar magnetic fields studied in this paper makes such calculations considerably less computationally demanding compared to modelling of stars with multi-kG fields. Additionally, the chemical spot maps of \aur\ and \uma\ can be deemed, on average, somewhat more robust than previous results obtained for strong-field MCP stars since in this case the abundance DI is essentially decoupled from the problem of magnetic mapping. In this situation even a complete neglect of the magnetic field would not lead to significant errors in the resulting chemical spot maps \citep{kochukhov:2017}.

It is interesting to note that the horizontal distribution of one particular element, oxygen, in \uma\ \citep{rice:1997} obeys the diffusion theory predictions and exhibits a well-defined ring-like overabundance structure at the magnetic equator. There is only one other known case -- O and C in the cool Ap star HR\,3831 \citep{kochukhov:2004e} -- where this behaviour is observed. On the other hand, \citet{rice:2004} reported a much more complex O map for \aur. These authors speculated that this might be due to a more complex, quadrupole-dominated field topology of that star. Our results do not confirm this suggestion. Although the surface field map appears to be more structured for \aur\ than for \uma, at least some of this difference can be attributed to a higher \vs\ and availability of Stokes $QU$ information for the former star. This complexity notwithstanding, the overall field of \aur\ is still predominantly dipolar. In particular, the field inclination map traces a single, uninterrupted ring (see Fig.~\ref{fig:abn_aur}), which is morphologically indistinguishable from the field inclination map of \uma\ (Fig.~\ref{fig:abn_uma}). This makes the dramatic difference between the O distributions of the two stars even more puzzling, again hinting that the local magnetic characteristics are not the only parameters governing the chemical structure formation.

In conclusion, we note that \uma\ is an MCP star with by far the largest angular diameter. This makes it the most promising target for direct interferometric star spot studies \citep{shulyak:2014a} similar to those recently carried out for cool active giants \citep{roettenbacher:2017}. Our work provides important constraints on the key stellar parameters and supplies reference surface map data for future interferometric studies of non-uniform brightness distributions associated with chemical spots. Such investigations are already feasible with existing facilities such as the CHARA array and VLTI. Future interferometers might even incorporate spectropolarimetry for spatially resolved magnetic field topology imaging \citep{rousselet-perraut:2000,rousselet-perraut:2004}. \uma\ and, to a lesser extent, \aur\ will be, no doubt, prime targets for this work.

\begin{acknowledgements}
This work is based on observations obtained at the Bernard Lyot Telescope (TBL, Pic du Midi, France) of the Midi-Pyr\'en\'ees Observatory, which is operated by the Institut National des Sciences de l'Univers of the Centre National de la Recherche Scientifique of France. Also based on observations obtained at the Canada-France-Hawaii Telescope (CFHT), which is operated by the National Research Council of Canada, the Institut National des Sciences de l'Univers of the Centre National de la Recherche Scientifique of France, and the University of Hawaii.
OK acknowledges support by the Swedish Research Council (project 621-2014-5720), and the Swedish National Space Board (projects 185/14, 137/17).
MS is supported by Natural  Sciences and Engineering Research Council (NSERC) of Canada Postdoctoral Fellowship program. 
\end{acknowledgements}

%\bibliographystyle{aa}
%\bibliography{astro_papers}

\begin{appendix}

\section{Observing logs and longitudinal field measurements}

\begin{table*}[!th]
\centering
\caption{Journal of spectropolarimetric observations of \aur\ and \uma. 
\label{tbl:obs}}
%{\small
\begin{tabular}{r l l l c c c}
\hline\hline
Star  & UT date    & HJD          & Phase & Stokes & $t_{\rm exp}$ (s) & SNR\\ 
\hline
\aur\ & 2006-12-01 & 2454070.8582 & 0.442 &   $V$ &           140 &            367  \\ %  0.0000 0.0000
      & 2006-12-05 & 2454074.9291 & 0.567 & $VQU$ &   140/140/140 & 1350/1369/1387  \\ %  0.0021 0.0021
      & 2006-12-06 & 2454076.1500 & 0.904 & $VQU$ &      52/52/52 &    805/804/804  \\ %  0.0016 0.0016
      & 2008-01-07 & 2454473.4221 & 0.688 & $VQU$ &   240/240/240 &    809/833/833  \\ %  0.0030 0.0030
      & 2008-01-09 & 2454475.2751 & 0.200 & $VQU$ &   240/240/240 &   881/901/1033  \\ %  0.0035 0.0035
      & 2008-01-23 & 2454489.0075 & 0.995 & $VQU$ &   240/240/240 & 1360/1472/1657  \\ %  0.0027 0.0027
      & 2008-01-24 & 2454490.0160 & 0.274 & $VQU$ &   160/160/160 & 1332/1359/1264  \\ %  0.0022 0.0022
      & 2008-01-26 & 2454491.7996 & 0.767 & $VQU$ &   160/160/160 & 1504/1497/1451  \\ %  0.0022 0.0022
      & 2008-01-30 & 2454495.8796 & 0.894 & $VQU$ &   160/160/160 &    905/827/823  \\ %  0.0022 0.0022
      & 2016-09-11 & 2457642.6659 & 0.493 & $VQU$ & 960/1920/1920 & 1009/2763/2533  \\ %  0.0152 0.0195
      & 2016-09-29 & 2457660.6162 & 0.454 & $VQU$ & 960/1920/1920 & 1563/2345/2396  \\ %  0.0161 0.0204
      & 2016-10-26 & 2457688.4827 & 0.155 & $VQU$ & 960/1920/1920 & 1620/1996/2252  \\ %  0.0151 0.0193
      & 2016-10-31 & 2457692.6777 & 0.314 & $VQU$ & 960/1920/1920 & 1640/2511/2322  \\ %  0.0152 0.0196
      & 2016-11-02 & 2457694.6299 & 0.853 & $VQU$ & 960/1920/1920 & 1799/2231/2069  \\ %  0.0151 0.0194
      & 2016-11-03 & 2457695.5669 & 0.112 & $VQU$ & 960/1920/1920 & 1936/2641/2710  \\ %  0.0153 0.0196
      & 2016-11-28 & 2457720.6806 & 0.052 & $VQU$ & 960/1920/1920 & 1660/2379/2329  \\ %  0.0154 0.0197
      & 2016-12-02 & 2457725.3928 & 0.354 & $VQU$ & 960/1920/1920 & 2047/2806/2773  \\ %  0.0154 0.0198
      & 2016-12-03 & 2457726.3710 & 0.625 & $VQU$ & 960/1920/1920 & 1724/1716/1893  \\ %  0.0151 0.0194
      & 2016-12-08 & 2457730.6432 & 0.805 & $VQU$ & 960/1920/1920 & 1836/2733/2596  \\ %  0.0151 0.0194
      & 2016-12-13 & 2457736.3999 & 0.396 & $VQU$ & 960/1920/1920 & 1402/1872/1986  \\ %  0.0151 0.0194
      & 2016-12-16 & 2457738.5158 & 0.981 & $VQU$ & 960/1920/1920 &  965/2149/1979  \\ %  0.0152 0.0195
      & 2017-01-31 & 2457785.4569 & 0.953 & $VQU$ & 960/1920/1920 & 1887/1897/2578  \\ %  0.0153 0.0197
      & 2017-03-02 & 2457815.4501 & 0.241 & $VQU$ & 960/1920/1920 & 1585/2112/2228  \\ %  0.0151 0.0194
      & 2017-03-14 & 2457827.4029 & 0.544 & $VQU$ & 960/1920/1920 & 1046/1457/1515  \\ %  0.0152 0.0195
      & 2017-03-19 & 2457832.4698 & 0.945 & $VQU$ & 480/1920/1920 & 1192/2699/2610  \\ %  0.0140 0.0172
      & 2017-03-29 & 2457842.4091 & 0.691 & $VQU$ & 960/1920/1920 & 1425/2416/2070  \\ %  0.0152 0.0196
      & 2017-04-06 & 2457850.3990 & 0.899 &  $QU$ &     1920/2400 &      1623/1944  \\ %  0.0097 0.0173
      & 2017-04-11 & 2457855.3712 & 0.273 & $VQU$ & 960/1920/1920 & 1943/2408/2480  \\ %  0.0151 0.0193
      & 2017-04-18 & 2457862.3690 & 0.207 & $VQU$ & 480/1920/1920 & 1188/2601/2613  \\ %  0.0140 0.0172
\hline
\uma\ & 2014-05-18 & 2456796.4729 & 0.121 &   $V$ &           128 &           1152  \\ %  0.0000 0.0000
 & 2016-12-10 & 2457732.7640 & 0.117 & $VQU$ &   156/312/312 &  906/1449/1267  \\ %  0.0029 0.0033
 & 2016-12-11 & 2457733.7380 & 0.309 & $VQU$ &   156/312/312 & 1050/1508/1465  \\ %  0.0029 0.0033
 & 2016-12-13 & 2457735.6402 & 0.683 & $VQU$ &   156/312/312 &  980/1533/1418  \\ %  0.0029 0.0033
 & 2016-12-14 & 2457736.6291 & 0.877 & $VQU$ &   156/312/312 &  680/1129/1104  \\ %  0.0029 0.0033
 & 2016-12-15 & 2457737.6246 & 0.073 & $VQU$ &   156/312/312 &    726/712/960  \\ %  0.0029 0.0033
 & 2017-01-07 & 2457760.5537 & 0.579 & $VQU$ &   156/312/312 & 1186/1710/1703  \\ %  0.0029 0.0033
 & 2017-01-08 & 2457761.5783 & 0.780 & $VQU$ &   156/312/312 &  993/1418/1428  \\ %  0.0029 0.0033
 & 2017-02-01 & 2457785.6625 & 0.513 & $VQU$ &   156/312/312 &    288/605/496  \\ %  0.0028 0.0032
 & 2017-02-14 & 2457799.4896 & 0.230 & $VQU$ &   156/312/312 &  919/1353/1310  \\ %  0.0029 0.0033
 & 2017-02-15 & 2457799.6860 & 0.269 & $VQU$ &   156/312/312 & 1039/1588/1505  \\ %  0.0029 0.0033
 & 2017-02-16 & 2457800.5101 & 0.431 & $VQU$ &   156/312/312 &  880/1293/1261  \\ %  0.0029 0.0033
 & 2017-02-16 & 2457800.7469 & 0.477 & $VQU$ &   156/312/312 & 1124/1601/1680  \\ %  0.0029 0.0033
 & 2017-02-17 & 2457801.5465 & 0.634 & $VQU$ &   156/312/312 & 1055/1504/1500  \\ %  0.0029 0.0033
 & 2017-02-17 & 2457801.6818 & 0.661 & $VQU$ &   156/312/312 & 1140/1640/1649  \\ %  0.0029 0.0033
 & 2017-02-18 & 2457802.5748 & 0.836 & $VQU$ &   156/312/312 & 1104/1322/1502  \\ %  0.0029 0.0033
 & 2017-02-18 & 2457802.6756 & 0.856 & $VQU$ &   156/312/312 & 1219/1618/1649  \\ %  0.0028 0.0032
 & 2017-02-19 & 2457803.5263 & 0.023 & $VQU$ &   156/312/312 & 1164/1703/1665  \\ %  0.0029 0.0033
 & 2017-02-20 & 2457804.5838 & 0.231 & $VQU$ &   156/312/312 &  930/1578/1518  \\ %  0.0029 0.0033
 & 2017-02-22 & 2457806.7073 & 0.649 & $VQU$ &   156/312/312 &  830/1216/1188  \\ %  0.0029 0.0033
 & 2017-02-22 & 2457807.4838 & 0.801 & $VQU$ &   156/312/312 &  622/1390/1353  \\ %  0.0029 0.0033
 & 2017-03-15 & 2457828.4415 & 0.920 & $VQU$ &   156/312/312 & 1218/1672/1730  \\ %  0.0029 0.0033
 & 2017-03-16 & 2457828.6590 & 0.962 & $VQU$ &   156/312/312 & 1251/1758/1670  \\ %  0.0029 0.0033
 & 2017-03-17 & 2457829.7183 & 0.171 & $VQU$ &   156/312/312 & 1193/1776/1759  \\ %  0.0029 0.0033
 & 2017-03-19 & 2457831.5499 & 0.531 & $VQU$ &   156/312/312 & 1197/1734/1701  \\ %  0.0029 0.0033
 & 2017-03-20 & 2457832.5345 & 0.724 & $VQU$ &   156/312/312 & 1185/1661/1698  \\ %  0.0029 0.0033
 & 2017-04-17 & 2457861.3410 & 0.385 & $VQU$ &   156/312/312 & 1248/1760/1754  \\ %  0.0029 0.0033
\hline
\end{tabular}
\tablefoot{The columns give the target name, the UT and heliocentric Julian dates at mid-observation, the corresponding rotational phase, the Stokes parameters obtained, the total exposure time and the resulting SNR. The latter refers to 1.8~\kms\ velocity bin of the extracted spectrum and was obtained from the median error bar in the 4500--6000~\AA\ wavelength region.}
%}
\end{table*}

\begin{table}[!th]
\centering
\caption{LSD and hydrogen line mean longitudinal magnetic field measurements of \aur\ and \uma. 
\label{tbl:bz}}
%{\small
\begin{tabular}{l l l r r}
\hline\hline
Star & HJD  & Phase & $\langle B_{\rm z} \rangle_{\rm LSD}$ (G) & $\langle B_{\rm z} \rangle_{\rm H}$ (G) \\ 
\hline
\aur\  & 2454070.8582 & 0.442 & $ 308\pm23$ & $ 225\pm88$ \\
 & 2454074.9252 & 0.566 & $ 258\pm 7$ & $ 225\pm27$ \\
 & 2454076.1472 & 0.903 & $ -89\pm13$ & $-137\pm44$ \\
 & 2454473.4275 & 0.690 & $ 218\pm12$ & $ 258\pm42$ \\
 & 2454475.2684 & 0.199 & $ -55\pm10$ & $ -64\pm35$ \\
 & 2454489.0026 & 0.994 & $-171\pm 8$ & $-169\pm26$ \\
 & 2454490.0121 & 0.273 & $  97\pm 7$ & $  54\pm23$ \\
 & 2454491.7956 & 0.766 & $ 150\pm 8$ & $ 114\pm25$ \\
 & 2454495.8756 & 0.893 & $ -64\pm12$ & $-105\pm36$ \\
 & 2457642.6921 & 0.501 & $ 272\pm 9$ & $ 270\pm35$ \\
 & 2457660.6443 & 0.462 & $ 275\pm 6$ & $ 285\pm23$ \\
 & 2457688.5087 & 0.162 & $-107\pm 6$ & $ -93\pm17$ \\
 & 2457692.7040 & 0.321 & $ 173\pm 5$ & $ 180\pm18$ \\
 & 2457694.6560 & 0.861 & $  11\pm 6$ & $ -32\pm17$ \\
 & 2457695.5932 & 0.120 & $-159\pm 5$ & $-151\pm14$ \\
 & 2457720.7071 & 0.060 & $-194\pm 6$ & $-166\pm18$ \\
 & 2457725.4192 & 0.362 & $ 233\pm 4$ & $ 191\pm15$ \\
 & 2457726.3970 & 0.632 & $ 238\pm 6$ & $ 185\pm17$ \\
 & 2457730.6693 & 0.813 & $  91\pm 6$ & $  36\pm16$ \\
 & 2457736.4261 & 0.403 & $ 258\pm 6$ & $ 222\pm19$ \\
 & 2457738.5420 & 0.988 & $-185\pm12$ & $-132\pm38$ \\
 & 2457785.4833 & 0.960 & $-133\pm 6$ & $-168\pm16$ \\
 & 2457815.4762 & 0.249 & $  44\pm 6$ & $  31\pm17$ \\
 & 2457827.4291 & 0.552 & $ 244\pm 8$ & $ 179\pm27$ \\
 & 2457832.4933 & 0.951 & $-135\pm 9$ & $-129\pm28$ \\
 & 2457842.4353 & 0.699 & $ 220\pm 7$ & $ 186\pm21$ \\
 & 2457855.3972 & 0.281 & $ 115\pm 5$ & $ 102\pm13$ \\
 & 2457862.3925 & 0.214 & $ -10\pm 8$ & $  11\pm24$ \\
\hline
\uma\ & 2456796.4729 & 0.121 & $  96\pm 4$ & $  54\pm19$ \\
 & 2457732.7710 & 0.119 & $  90\pm 5$ & $  71\pm21$ \\
 & 2457733.7451 & 0.310 & $  17\pm 5$ & $  27\pm18$ \\
 & 2457735.6473 & 0.684 & $ -14\pm 6$ & $ -31\pm22$ \\
 & 2457736.6361 & 0.878 & $  89\pm 7$ & $  84\pm30$ \\
 & 2457737.6317 & 0.074 & $  96\pm 6$ & $  14\pm29$ \\
 & 2457760.5608 & 0.580 & $ -55\pm 5$ & $  -6\pm17$ \\
 & 2457761.5853 & 0.781 & $  41\pm 6$ & $  30\pm22$ \\
 & 2457785.6694 & 0.514 & $ -54\pm19$ & $  -3\pm93$ \\
 & 2457799.4966 & 0.232 & $  58\pm 6$ & $  48\pm19$ \\
 & 2457799.6929 & 0.270 & $  38\pm 5$ & $  32\pm18$ \\
 & 2457800.5171 & 0.432 & $ -46\pm 6$ & $ -34\pm23$ \\
 & 2457800.7540 & 0.479 & $ -67\pm 5$ & $ -37\pm18$ \\
 & 2457801.5535 & 0.636 & $ -41\pm 5$ & $ -28\pm21$ \\
 & 2457801.6889 & 0.662 & $ -27\pm 5$ & $ -15\pm16$ \\
 & 2457802.5818 & 0.838 & $  77\pm 5$ & $  40\pm17$ \\
 & 2457802.6825 & 0.858 & $  82\pm 4$ & $  42\pm16$ \\
 & 2457803.5333 & 0.025 & $  92\pm 4$ & $  42\pm18$ \\
 & 2457804.5908 & 0.233 & $  57\pm 6$ & $  50\pm23$ \\
 & 2457806.7144 & 0.650 & $ -44\pm 7$ & $ -15\pm20$ \\
 & 2457807.4907 & 0.803 & $  67\pm 8$ & $  42\pm36$ \\
 & 2457828.4345 & 0.918 & $  95\pm 4$ & $  69\pm16$ \\
 & 2457828.6520 & 0.961 & $  93\pm 4$ & $  98\pm11$ \\
 & 2457829.7113 & 0.169 & $  88\pm 4$ & $  20\pm16$ \\
 & 2457831.5428 & 0.529 & $ -65\pm 5$ & $  -7\pm17$ \\
 & 2457832.5274 & 0.723 & $  17\pm 5$ & $  63\pm16$ \\
 & 2457861.3340 & 0.384 & $ -27\pm 5$ & $ -46\pm18$ \\
\hline
\end{tabular}
%}
\end{table}

\clearpage 

\section{Comparison of $IQUV$ and $IV$ ZDI inversions for \aur}
\label{zdi_test}

The two Ap stars studied in our paper were analysed using a somewhat different ZDI methodology. In one case (\aur), we were able to use all four Stokes parameter spectra for magnetic inversions. For the other star (\uma), only the Stokes $IV$ data were available for modelling. In addition, tomographic mapping of \aur\ benefited from a higher \vs\ of that star. It is of interest to assess the influence of these differences on the reconstructed surface structure maps.

In the past, several studies compared inversions using full Stokes vector observations and $IV$ spectra \citep{piskunov:2002a,kochukhov:2002c,kochukhov:2010,kochukhov:2016a,rosen:2015}. The impact of the limited information content of circular polarisation spectra varied from moderate to severe, depending on the stellar parameters, degree of field complexity, and the field strength. Considering these results, it is pertinent to perform dedicated simulations to study the influence of input data and stellar parameters on the inversion results for the two MCP stars targeted by our study.

To this end, we carried out a test Stokes $IV$ ZDI inversion for \aur\ using LSD spectra calculated for the Cr map and magnetic field geometry obtained in Sect.~\ref{zdi_aur}, the same set of rotation phases as in the actual observations and using $i=123\fdg7$, but adopting \vs\,=\,35~\kms, similar to the projected rotational velocity of \uma. Random noise was added according to the SNR of observations at specific rotation phases. These simulated data were used for the simultaneous ZDI mapping of the magnetic field structure and Cr abundance distribution starting from the zero magnetic field and a uniform abundance map initial guesses.

\begin{figure*}[!h]
\centering
\fups{15cm}{0}{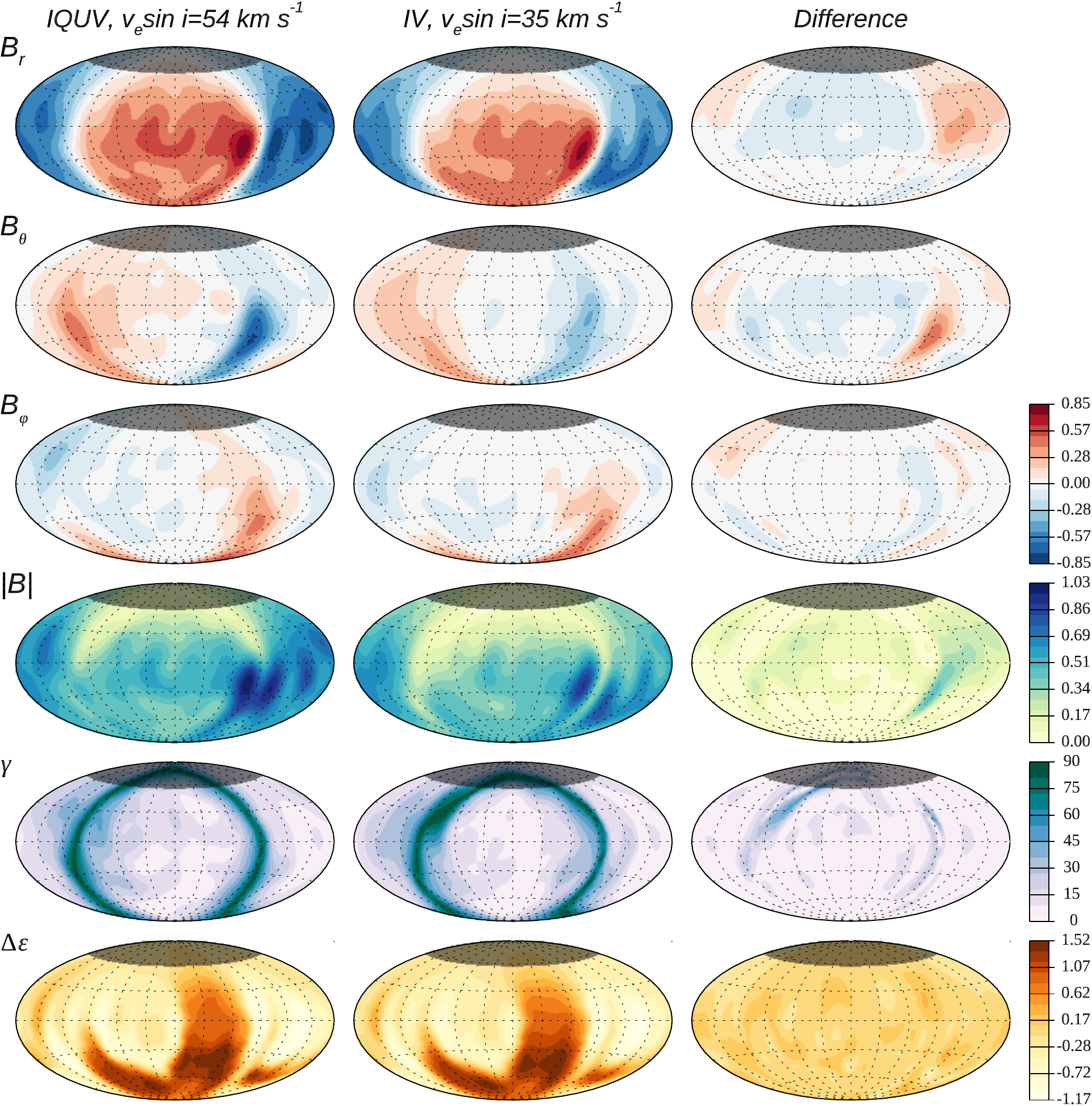}
\caption{Comparison of the Cr four Stokes parameter ZDI results for \aur\ (left column) with the Stokes $IV$ inversion (middle column) from the spectra simulated for \vs\,=\,35~\kms\ and the same surface abundance and magnetic field distributions. The right column shows the difference between the two sets of maps. The rows correspond to, from top to bottom, the radial, meridional, and azimuthal magnetic field components, the field modulus, the field inclination angle, and the relative Cr abundance. The side colour bars give the scale in kG for the magnetic field maps, in degrees for the field inclination, and in $\Delta$$\log (N_{\rm Cr}/N_{\rm tot})$ units for the Cr abundance distribution. The shaded part of stellar surface in the upper part of each map is invisible to the observer.}
\label{fig:zdi_comp}
\end{figure*}

The outcome of this numerical experiment is presented in Fig.~\ref{fig:zdi_comp}. We show maps in the Hammer-Aitoff projection for the three magnetic field vector components, the field modulus, the local field vector inclination, and Cr abundance. The two sets of maps compare the outcome of the Stokes $IQUV$ inversion and the Stokes $IV$ reconstruction from the simulated data. The difference maps are presented as well.

We found that neglecting the Stokes $QU$ data and reducing \vs\ has a relatively small impact on the ZDI inversion. Some minor magnetic surface structure details appear to be smoothed out in the distributions obtained from Stokes $IV$ spectra. The field strength map is also less structured. On the other hand, the maps of the field inclination and Cr abundance obtained in the two inversions appear nearly identical in Fig.~\ref{fig:zdi_comp}, suggesting that reconstruction of these quantities is particularly robust against the loss of information in the Stokes $QU$ spectra.

The mean absolute difference between the Stokes $IQUV$ and $IV$ reconstruction results amounts to 86~G for the radial field component, 64~G for the meridional field, 40~G for the azimuthal field, and 94~G for the field modulus. This corresponds to about 10\% of the peak values in the respective surface maps. The average difference of the field inclination reconstruction is $7\fdg4$. The Cr abundance maps agree to within 0.07~dex. The global magnetic field characteristics do not change significantly between the full Stokes vector mapping and the Stokes $IV$ reconstruction. For the latter case we infer that 76.0\% of the magnetic energy is contained in the $\ell=1$ harmonic component and 82.8\% of the energy is concentrated in the poloidal field. These numbers are within 2.5\% of the original Stokes $IQUV$ inversion results (see Table~\ref{tbl:zdi}).

We conclude that the inclusion of Stokes $QU$ observations in ZDI inversions and a higher \vs\ of \aur\ relative to \uma\ has a small impact on the quality of inferred magnetic maps and is entirely negligible for reconstruction of the field inclination and element abundance distributions. However, we caution that these conclusions should not be extrapolated to magnetic stars with parameters significantly different from \aur\ or \uma. For example, the inclusion of $QU$ spectropolarimetric data is likely to have a larger impact on ZDI analyses of MCP stars with stronger and/or more complex surface fields and on studies of stars rotating significantly more slowly than the two objects investigated here.

\end{appendix}

\end{document}